%% file: ProgressivePOD.tex
\documentclass[10pt]{article}
\usepackage{color}
\usepackage{geometry}
\usepackage[algo2e,ruled,lined,titlenumbered]{algorithm2e}

\include{notations}

\usepackage{amssymb} 
\usepackage{amsmath} 
\usepackage{stmaryrd} 
\usepackage{bm}
\usepackage{graphicx}
\usepackage{comment}
\newcommand{\argmax}{\operatornamewithlimits{argmax}}

\usepackage{algorithmic,algorithm}

\title{Statistical extraction of process zones and representative subspaces in fracture of random composites}

\author{P. Kerfriden$^{1}$\footnote{mailing adress: Cardiff University School of Engineering, Queen's building, The Parade, CF243AA Cardiff, Wales, United Kingdom ; email: kerfridenp@cardiff.ac.uk} , K.M. Schmidt$^{2}$, T. Rabczuk$^{3}$ and S.P.A. Bordas$^{1}$
\\ \\
$\begin{array}{cl}
^{1} & \textrm{Cardiff University, School of Engineering} \\
& \textrm{Queen's Buildings, The Parade, Cardiff CF24 3AA, Wales, UK} \\
^{2} & \textrm{Cardiff University, School of Mathematics} \\
& \textrm{Senghennydd Road, Cardiff CF24 4AG, Wales, UK} \\
^{3} & \textrm{Institute of Structural Mechanics, Bauhaus-University Weimar} \\
& \textrm{Marienstra\ss e 15, 99423 Weimar, Germany}
\end{array}$
}

\begin{document}
\maketitle

\begin{abstract}
We propose to identify process zones in heterogeneous materials by tailored statistical tools. The process zone is redefined as the part of the structure where the random process cannot be correctly approximated in a low-dimensional deterministic space. Such a low-dimensional space is obtained by a spectral analysis performed on pre-computed solution samples. A greedy algorithm is proposed to identify both process zone and low-dimensional representative subspace for the solution in the complementary region. In addition to the novelty of the tools proposed in this paper for the analysis of localised phenomena, we show that the reduced space generated by the method is a valid basis for the construction of a reduced order model.
\end{abstract}



\section{Introduction}



\input{Introduction}


\section{Lattice model of fracture in particulate composites}


\input{modele}

\subsection{Solution strategy}

\input{solution}


\section{Extraction of coherent structures by the proper orthogonal decomposition}


\input{POD}


\input{results}


\section{Conclusion}

We proposed a method based on the statistical analysis of solution samples for extracting the process zone of a problem of fracture in random heterogeneous media. We showed that the definition of the process zone is not unique, but is in fact parametrised by the level of non-correlation of the random process which is obtained in the complementary part of the domain. The method generates a reduced space for this complementary part, in which the solution to the problem of fracture in random composite is optimally approximated. We showed the potential of this novel strategy as a tool for the analysis of localised physical phenomena and as a means to speed up parametric problems via preconditioning or reduced order modelling.

\section{Acknowledgements}

The authors acknowledge the financial support of the Royal Academy of Engineering and of the Leverhulme Trust for Bordas' Senior Research Fellowship ``Towards the next generation surgical simulators'' (2009-2010) as well as the support of EPSRC under grant EP/G042705/1 Increased Reliability for Industrially Relevant Automatic Crack Growth Simulation with the eXtended Finite Element Method. Mrs Susanne Claus (Cardiff University School of Mathematics) is also thanked for helping proofread this article.

\bibliographystyle{unsrt}
\bibliography{bibliography}

\end{document}

%% file: notations.tex
\newcommand{\V}[1]{\underline{#1}}
\newcommand{\M}[1]{\underline{\underline{#1}}}

\newcommand{\disp}{\V{u}}

\newcommand{\disptest}{\V{u}^{\star}}

\newcommand{\domega}{\partial \Omega}
\newcommand{\domegau}{\partial \Omega_u}
\newcommand{\domegaf}{\partial \Omega_f}
\newcommand{\ud}{\V{U}_d}
\newcommand{\fd}{\V{f}_d}
\newcommand{\Fd}{\V{F}_d}
\newcommand{\stress}{\M{\sigma}}

\newcommand{\straintest}{\M{\epsilon}(\disptest)}

\newcommand{\dispdiscr}{\V{\mathbf{U}}}

\newcommand{\fint}{\V{\mathbf{F}}_{\textrm{Int}}}
\newcommand{\fext}{\V{\mathbf{F}}_{\textrm{Ext}}}
\newcommand{\zerodiscr}{\V{\mathbf{0}}}

\newcommand{\X}{\V{\mathbf{X}}}

\newcommand{\basereduc}{\M{\boldsymbol{\Phi}}}
\newcommand{\coeffreduc}{\V{\boldsymbol{\alpha}}}

%% file: Introduction.tex
In order to predict complex physical phenomena, one has to devise models which describe these phenomena at the appropriate, sufficiently fine scale. In the case of fracture in heterogeneous structures, it is usually observed that models need to be established at the scale of these heterogeneities. For composite laminates for instance, predicting fracture requires an explicit description of the damage process within each individual ply, and of the thin interface transition zone between two adjacent plies \cite{allixleveque1998,ladevezelubineau2002,pinhoIannucci2006}. Similarly, fracture models for concrete need to take into account explicitly the interface transition zone between cement and aggregates \cite{vanmiervanvliet2002,karihalooshao2003,scrivenercrumbie2004}. Models that are built on this observation describe fracture at a meso-scale (i.e.: scale of the mesoconstituents). With today's ever increasing computing power available to engineering, the simulation of large engineering components using such mesoscale models is within reach, albeit at a considerable cost. However, in engineering design processes, a prohibitively high number of such simulations may be necessary. One might be interested in the sensibility of the fracture process to some design parameters, or in the effect of observed variabilities on the strength of the structure. Consequently, constructing reduced order models based on a mesoscale representation of fracture is an issue of tremendous importance in today's computational mechanics.

Two families of approaches address this problematic from a different angle. Homogenisation-based reduced order modelling proposes to use the fact that the solution to multiscale problems can be split into two additive contributions: a macroscale contribution, which is smooth and varies slowly in the structure, and its microscale counterpart which varies with the fine-scale structural heterogeneities. Under the hypothesis of periodicity or ergodicity of the heterogeneities, one can establish a formal separation of scale \cite{sanchez-palencia1980,suquet1987}: the macroscale part of the solution is obtained by solving a macroscale problem at the scale of the structure, while the macroscale problem itself is obtained by solving microscale problems on small subdomains of the structure over which the effect of the periodic heterogeneities can be averaged. In the case of fracture in heterogeneous media, the difficulty is that the assumptions of periodicity or ergodicity are violated in the damaged region, which jeopardises the accuracy of this class of methods. Filtering out the effect of the crack to retrieve this periodicity is currently an active area of research \cite{massartpeerlings2007,belytschkoloehnert2007,allixkerfriden2010b,nguyenlloberas-valls2012,coenenkouznetsova2012}. Algebra-based reduced order models follow a different approach. The idea is to define a subspace of low dimension in which the fine-scale solution to the problem of interest is well-approximated. The fine-scale problem is then projected onto this subspace, which yields a problem of small dimension, which takes into account the fine-scale features of the problem. One of the major difficulties here is of course the identification this subspace.

Algebra-based reduced order modelling has been extensively studied in the literature in the case of linear problems. The extraction of invariant representative subspaces based on the spectral analysis of linear problem is the basis for a large class of methods, which comprises modal synthesis, balanced truncation and moment matching. The construction of reduced order model based on such eigenanalysis is relatively well-established (we refer to the review in \cite{antoulassorensen2001}). In the case of nonlinear problems, some breakthrough has been made over the last decade, with the development of reduced order modelling techniques based on the proper orthogonal decomposition \cite{ravindran2000,kunishvolkwein2003,legresleyalonso2003,astridweiland2008,carlbergbou-mosleh2011,kerfridengosselet2010}, on the reduced basis method \cite{barraultmaday2004}, or on the \textit{a priori} hyperreduction method \cite{ryckelynck2008,kerfridengosselet2010}. Generally speaking, the idea is to extend the concept of extraction of spectral invariants of the problem by considering a subset of particular solutions to the fine-scale nonlinear problem, called ``snapshot''. This snapshot, in the case of parametric analysis, is a set of solutions corresponding to particular values of the parameters. In the case of time-dependent nonlinear problems, the fine-scale solutions to the first time-steps can be used to identify a representative subspace, which in turn serves for the construction of a reduced order model for the cheap solution in subsequent time steps.

 \begin{figure*}[htb]
        \centering
        \includegraphics[width=0.9 \linewidth]{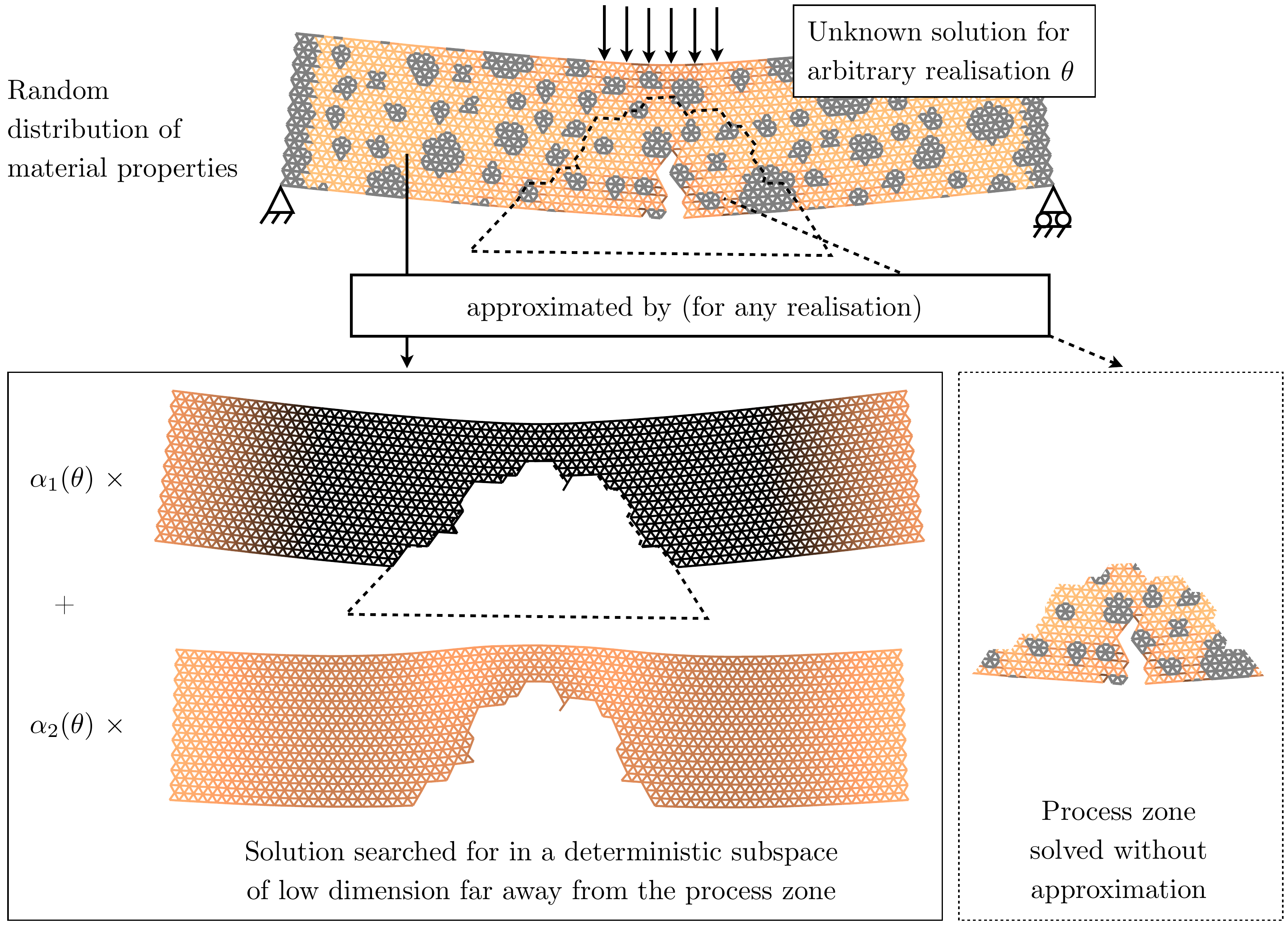}
        \caption{Principle of the reduced order modelling for problems with localised random effects while excluding regions of the domain where local lack of correlation is observed. The proposed work aims at providing an objective methodology to identify these regions and construct a representative reduced space for the solution corresponding to the complementary ``smooth'' domain.}
        \label{fig:Principle}
 \end{figure*}

However, most of the successful applications of algebra-based nonlinear reduced order modelling have been dedicated to mildly nonlinear problems, or more precisely to problems that exhibit a mild nonlinear dependency of the fine-scale solution on time or parameters. Fracture mechanics is characterised by strong nonlinearities in the region where cracks initiate and propagate, which jeopardises the identification of representative low-dimensional subspaces (this fact will be discussed in more details in the core of the paper). Attempts to exclude process zones (i.e.: the zones where mechanical energy is dissipated by damage mechanisms) from the reduction process have been proposed, allowing for the construction a reduced order model far away from the source of nonlinearity, while solving in the process zone without approximation \cite{haryadikapania1998,kerfridenpassieux2011} (a schematic of this general idea applied to the case of fracture of random composites is given in figure \ref{fig:Principle}). Similar strategies have been developed in other contexts where reduced order modelling cannot be efficiently applied in a particular region (usually the region of interest) of the structure, using for instance substructuring \cite{barbonegivoli2003,rickeltreese2006,buffonitelib2009,kerfridengoury2012} or local enrichment \cite{legresleyalonso2003,ammar2011}. So far, the identification of these process zones has been made \textit{a priori}, from physical or empirical observations. The present work is an attempt to provide some objectivity in the determination of the zones where algebra-based coarsening should not be performed.

The method proposed in this paper relies on a very simple idea. The process zone can simply be redefined as the part of the domain where reduced order modelling does not provide a satisfactory level of accuracy. Let us elaborate on the implications of this observation in the context of fracture of random materials. The problem of interest is depicted in figure \ref{fig:Principle}. The distribution of inclusions in a particulate composite follows a given probability law. We are interested in finding a low-dimensional subspace in which the solutions corresponding to all possible realisations of the particle distribution are well approximated. We use the classical idea of the snapshot proper orthogonal decomposition \cite{sirovich1987} to obtain such a subspace. One first computes explicitly the solution corresponding to a few particular realisations (the snapshot) of the random particle distribution. The reduced space is then defined as the subspace of the space spanned by the snapshots in which the projection of the snapshots is the closest from the original sample. Mathematically, the reduced space is obtained by a spectral analysis of the space spanned by the sample solutions. The problem is said to be reducible if such a space of low dimension indeed allows for a sufficiently small error of projection.  Fracture of random materials, as will be shown in this work, is not directly reducible in this sense. However, if one excludes the process zone from the spectral analysis, an acceptable level of reducibility is obtained in the remainder of the domain. Therefore, we can use the observation made previously and define the process zone as a region where the error of projection is too large. We keep using the term ``process zone'' classically used in fracture mechanics because the region associated to large local errors of projection is observed to correspond to zones where damage might occur in a statistical sense. But we emphasize the fact that the usual definition of the ``process zone'' is not strictly equivalent to the algorithmic redefinition used in this paper. To summarise the idea introduced in this paragraph, we look for a reduced space and, at the same time, for the associated domain in which the approximation provided by reduced order modelling is valid. 

Mathematically, this problem can be formulated as a problem of minimisation of the error of projection in the reduced space, the unknowns being the reduced space itself and the process zone. We propose to solve it by a greedy-type algorithm. One first perform a classical proper orthogonal decomposition of the snapshot where a fixed iterate of the process zone is excluded. In a second step, one looks for a small update of the process zone which minimises the error of projection, given the reduced space. We arrive at a sub-optimal solution of what we call the restricted POD.

Using the proposed tool, we find that the approximation error associated with a low-dimensional reduced space and a relatively confined process zone is acceptable for engineering applications. In addition, the dimensionality of the solution space, far away from the process zone, is relatively well-defined: one can easily identify the (small) dimension of the reduced space which contains meaningful information. However, the process zone itself is not well-defined. Indeed, the influence of the random cracks is felt far away from the physical discontinuities. As a consequence, the error associated with the restricted POD decreases at an almost constant rate with the size of the process zone. We conclude that there is not one definite process zone, but several valid process zones, each defined by the level of error of the reduced order model built in the complementary region.

The outline of the paper is the following. We describe in Section 2 the damageable lattice model that is used to simulate fracture in particulate composites (see some of the related work in \cite{vanmiervanvliet2002,lilliuvanmier2003,rabczukkim2004,karihalooshao2003,grasslbazant2009}). We briefly describe the random distribution of material properties, and set up the problem of fracture in particulate composites. The classical POD and Snapshot POD are presented in Section 3, along with cross-validation statistical error estimates \cite{bradleygong1983,jackson1993,cangelosigoriely2007}. We show numerically that the problem of interest is locally uncorrelated and review classical solutions to address this type of issues. In section 4, we define an extension of these approaches, the restricted POD. A progressive greedy-type algorithm meant to find an optimal decomposition associated with a spatial domain of validity is described next. We show that the proposed tool is relevant to the problem of fracture of random particulate composites in section 5, and discuss the possible improvements and applications in section 6.

%% file: modele.tex
\subsection{Lattice structure}

\begin{figure}[htb]
       \centering
       \includegraphics[width=0.75 \linewidth]{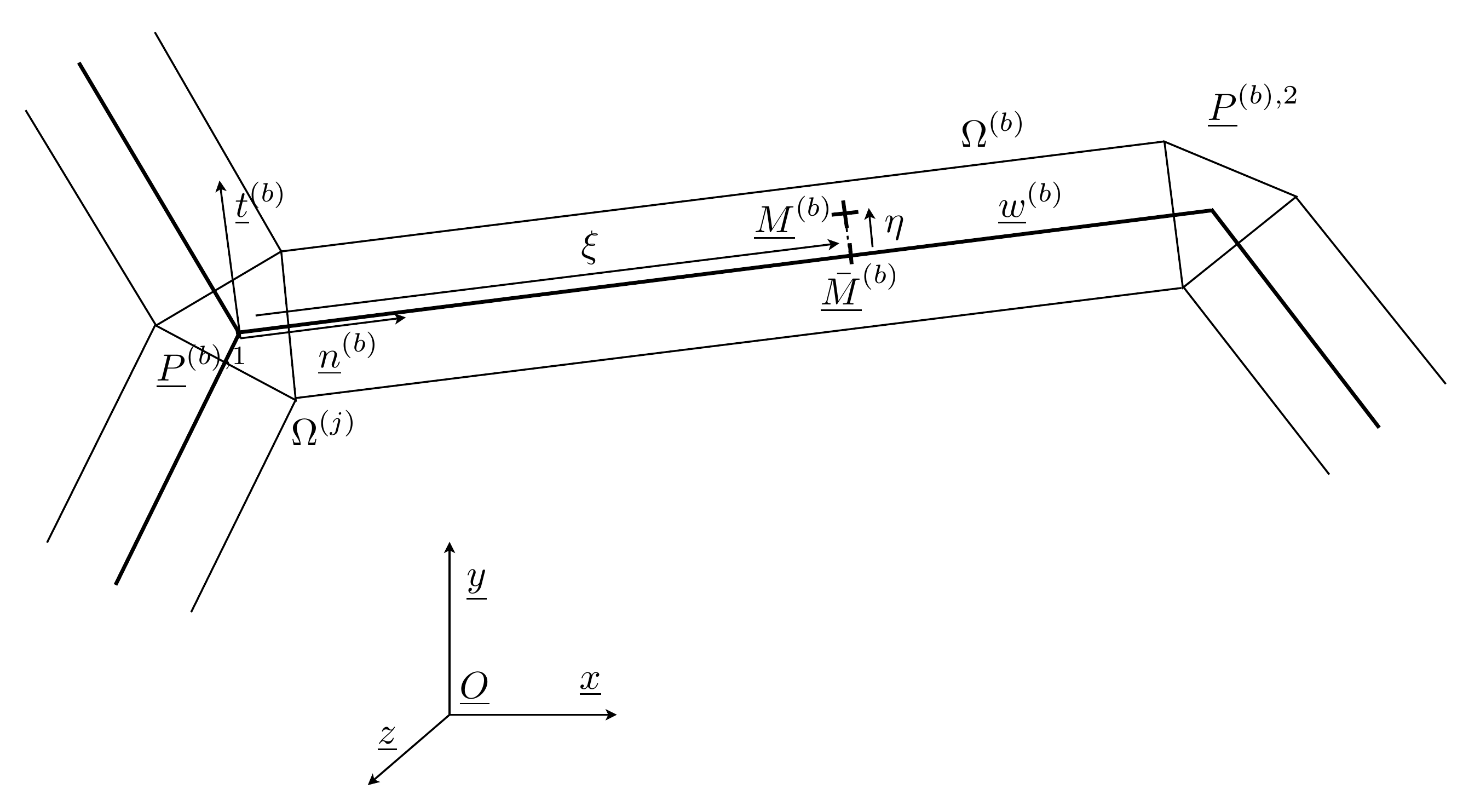}
       \caption{Definition of the beam network and its associated lattice model}
       \label{fig:LatticeDefinition}
\end{figure}

We consider a two-dimensional lattice occupying a continuous domain $\Omega$ with boundary $\domega$. Each of the $n_b$ lattice beams occupies domain $\Omega^{(b)}$. Sets of adjacent beams are linked by rigid joints occupying domains $\Omega^{(j)}$ such that $\Omega = \bigcup_{b \in \llbracket 1 , n_b \rrbracket} \Omega^{(b)} \cup \bigcup_{j \in \llbracket n_b+1 , n_b+n_j \rrbracket} \Omega^{(j)}$. $n_j$ is the number of joints. The structure is subjected to prescribed displacements $\ud$ on the part of its boundary denoted by $\domegau$, to prescribed tractions $\Fd$ on the complementary boundary $\domegaf = \domega \backslash \domegau$, and to a distributed body force $\fd$, over time interval $\mathcal{T}=[0,T]$. These boundary conditions depend on time, which will not be written explicitly, unless necessary. The evolution is supposed quasi-static, isothermal and we make the assumption of small perturbations (small displacements and strain). Let us define a global reference frame $\mathcal{R}=(\V{O},\V{x},\V{y},\V{z})$, where $\mathcal{B}=(\V{x},\V{y},\V{z})$ is an orthonormal basis and $\V{z}$ is orthogonal to the two-dimensional plane. Depending on the context, we will alternatively use the notations $\mathcal{R}=(\V{O},\V{x},\V{y})$ and $\mathcal{B}=(\V{x},\V{y})$.

We consider the limit case where each beam is infinitely slender (in the two-dimensional plane), and the domain occupied by the joints is of measure null. We therefore work with a network of unidimensional domains $(\omega^{(b)})_{b \in \llbracket 1 , n_b \rrbracket}$, such that $\omega = \bigcup_{b \in \llbracket 1 , n_b \rrbracket} \omega^{(b)}$ is embedded in $\mathbb{R}^2$, joining a set of grid points denoted by $\mathcal{P} = \{\V{P} \}_{i \in \llbracket 1, n_p\rrbracket}$. Let $\V{P}^{(b),1} \in \mathcal{P}$ and $\V{P}^{(b),2} \in \mathcal{P}$ be the two extremities of beam $(b)$. Line $(\V{P}^{(b),1},\V{P}^{(b),2})$ defines the neutral fibre of beam $b$. $\V{P}^{(b),1}$ and $\V{P}^{(b),2}$ are ordered such that if we define global indices $i$ and $j$ by $\V{P}_i = \V{P}^{(b),1}$ and $\V{P}_j = \V{P}^{(b),2}$, we enforce the condition $i<j$. We define the unit neutral axis vector of beam $b$ by $\V{n}^{(b)}=(\V{P}^{(b),2} - \V{P}^{(b),1}) / L^{(b)}$, where $ L^{(b)} = \| \V{P}^{(b),2} - \V{P}^{(b),1} \|_2$ is the length of beam $b$. The definition of a local reference frame attached to $b$ is done as follows: $\mathcal{R}^{(b)} = (\V{P}^{(b),1},\V{n}^{(b)},\V{t}^{(b)},\V{z})$ with $\V{t}^{(b)} = \V{z} \wedge \V{n}^{(b)}$. The local orthonormal basis is denoted by $\mathcal{B}^{(b)} = (\V{n}^{(b)},\V{t}^{(b)},\V{z})$. We will write the following coordinates of arbitrary point $\V{M}^{(b)}$ of $(b)$ in the local reference frame:
\begin{equation}
\V{P^{(b),1}M^{(b)}} = \xi \, \V{n}^{(b)} + \eta \, \V{t}^{(b)} + z \, \V{z} \, .
\end{equation}
The orthogonal projection of $\V{M}^{(b)}$ onto semi-segment $(\V{P}^{(b),1},\V{n}^{(b)})$ is denoted by $\V{\bar{M}}^{(b)}$ and is the point of the neutral fibre at linear coordinate $\xi$.
The cross sections (section orthogonal to the neutral fibre) of beam $(b)$ at points $\V{P}^{(b),1}$ and $\V{P}^{(b),2}$ will be respectively denoted by $S^{(b),1}$ and $S^{(b),2}$. A cross section at any other point $\V{\bar{M}}^{(b)}(\xi)$ will be $S^{(b)}(\xi)$.

Let us also define the subset of lattice grid points connecting at least two beams $\mathcal{J} = \{ P_i \}_{i \in \llbracket 1, n_j\rrbracket} \subset \mathcal{P}$ (centroids of joints ${(j)}$), the set of grid points at which Dirichlet boundary conditions are applied $\mathcal{P}_U = \{ P_i \}_{i \in \llbracket 1, n_u \rrbracket} \subset \mathcal{P}$ (centroids of beam cross sections that belong to $\partial \Omega_u$, and centroids of a joints whose boundaries intersect $\partial \Omega_u$) and the set of grid points at which Neumann boundary conditions are applied $\mathcal{P}_F = \{ P_i \}_{i \in \llbracket 1, n_f\rrbracket} \subset \mathcal{P}$ such that $\mathcal{P}_F \cup \mathcal{P}_U \cup \mathcal{J} = \mathcal{P}$ and $\mathcal{P}_F \cap \mathcal{P}_U = \{ \}$.

\subsubsection{Euler-Bernoulli approximation}

Let $\disp^{(b)}$ be the unknown displacement field in beam $(b)$, which belongs to the space $\mathcal{U}^{(b)}$ of kinematically admissible fields defined over $\Omega^{(b)}$:
\begin{equation}
\label{eq:kinematic}
\mathcal{U}^{{(b)}}=\left\{\disp^{(b)} \in H^1(\Omega^{(b)}) \ | \  \disp_{| \domegau} = \ud\right\} \, ,
\end{equation}
Let $\mathcal{U}^0$ be the associated vector space.
Under the assumptions made previously, the weak form of the balance equations reads, at any time $t \in [0,T]$:
\begin{equation}
\label{eq:equilibrium}
\begin{array}{l}
\displaystyle \text{find} \ \disp^{(b)} \in \mathcal{U}^{(b)} \ \text{such that:} \ \forall \disptest \in \mathcal{U}^{(b),0},  
\\ \displaystyle
\int_{\Omega^{(b)}} \stress^{(b)} : \straintest \, d \Omega = \int_{\Omega^{(b)}} \fd \cdot \disptest \, d \Omega 
+ \int_{ \partial \Omega^{(b)} \cap \domegaf } \Fd \cdot \disptest \, d \Gamma  
 + \int_{\partial \Omega^{(b)} \cap \partial \Omega^{(j)}} \stress^{(b)} \cdot \V{\tilde{n}}^{(b)} \cdot \disptest \, d \Gamma  \, ,
\end{array} 
\end{equation}
where $\stress^{(b)}$ is the Cauchy stress tensor and $\M{\epsilon}(\disp^{(b)}) = 1/2 \, (\M{\nabla} (\disp^{(b)}) + \M{\nabla} (\disp^{(b)})^T )$ is the symmetric part of the displacement gradient, and $\V{\tilde{n}}^{(b)} = \pm \V{n}^{(b)}$ is the outer normal to the beam. The last term in equation \eqref{eq:equilibrium} accounts for the reaction forces from the adjacent beams.
This balance equation needs to be complemented by a constitutive law between $\stress$ and $\M{\epsilon}$ (damage in our case), which will be detailed later on.

Let us now define the classical beam approximation of the previous problem on $\Omega^{(b)}$. The displacement $\disp^{(b)}$ is searched in a subspace $\mathcal{U}^{l,(b)}$ of $\mathcal{U}^{(b)}$ which satisfies the classical Euler-Bernoulli assumptions:
\begin{equation}
\label{eq:kinematic_lattice}
\displaystyle \mathcal{U}^{l,(b)}= 
\left\{ \disp^{(b)} \in H^1(\Omega^{(b)}) \ | 
\right. 
 \disp^{(b)} (\xi , \eta) 
= 
\begin{pmatrix}
v^{(b)}(\xi) - \theta^{(b)}(\xi) \, \eta \\
w^{(b)}(\xi)
\end{pmatrix}_{\mathcal{B}^{(b)}} 
\left. \displaystyle
 \textrm{and} \ w^{(b)}_{, \xi}(\xi) = \theta^{(b)}(\xi) \right\} \, ,
\end{equation}
Equation \eqref{eq:kinematic_lattice} defines a rigid body kinematic of each of the cross sections of the beam. The displacement is therefore uniquely defined by three unidimensional functions $v$, $w$ and $\theta$ defined on $\omega^{(b)}$. We will write $\V{q}^{(b)} = v^{(b)} \,  \V{n}^{(b)} +  w^{(b)} \, \V{t}^{(b)}$ the displacement of a point of the neutral axis and the local solution $S^{(b)} = (\V{q}^{(b)},\theta^{(b)}) $, searched in space $\mathcal{Q}^{(b)}$ defined by:
\begin{equation}
\displaystyle 
\mathcal{Q}^{(b)} = \left\{S^{(b)} = (\V{q}^{(b)},\theta^{(b)}) \ | \ \V{q}^{(b)} \in H^1(\omega^{(b)}) \, , 
 \theta^{(b)} \in H^1(\omega^{(b)}) \, , \,  
   \V{q}^{(b)}_{| \mathcal{P}_u} = \V{q}_d  \, ,  \,  \theta_{| \mathcal{P}_u} = \theta_d \right\} \, .
\end{equation}

Injecting the lattice approximation of the displacement into the balance equation \eqref{eq:equilibrium}, we obtain the following homogenised weak formulation:
\begin{equation}
\begin{array}{l}
\displaystyle \text{Find} \ S^{(b)} \in \mathcal{Q}^{(b)} \ \text{such that:} \  \forall \, S^{(b) \star} \in \mathcal{Q}^{(b),0},  
\\ \displaystyle
 \int_{\omega^{(b)}} \bar{\V{\sigma}}^{(b)} \cdot  \bar{\V{\epsilon}}^{(b) \star} \, d \xi = \int_{\omega^{(b)}} \bar{\V{f}}_d . \V{q}^\star \, d \xi  + \int_{\omega^{(b)}} m_d \, \theta^\star \, d \xi   
+ \sum_{P^{(b),i} \in \mathcal{P}_f} (\bar{\V{F}}_d \cdot \V{q}^\star)_{| P^{(b),i}} 
 \\ \displaystyle \qquad
 + \sum_{P^{(b),i} \in \mathcal{P}_f} (M_d \, \theta^\star)_{| P^{(b),i}}
+ \sum_{P^{(b),i} \in \mathcal{J}} (\bar{\V{F}}^{(b)} \cdot \V{q}^\star)_{| P^{(b),i}}
+ \sum_{P^{(b),i} \in \mathcal{J}} (M^{(b)} \, \theta^\star )_{| P^{(b),i}}
\end{array}
\end{equation}
The generalized strain in equation \eqref{eq:EquilibriumLatice} is:
\begin{equation}
\bar{\V{\epsilon}}^{(b)} (\xi)  =
\begin{pmatrix}
v^{(b)}_{,\xi} \\
\theta^{(b)}_{,\xi} 
\end{pmatrix}
=
\begin{pmatrix}
\V{q}^{(b)}_{,\xi} \cdot \V{n}^{(b)}\\
\theta^{(b)}_{,\xi} 
\end{pmatrix} \, ,
\end{equation}
where the generalized stress vector (axial force and moment) is defined as:
\begin{equation}
\bar{\V{\sigma}}^{(b)} (\xi) = 
\begin{pmatrix}
N^{(b)}(\xi) = \displaystyle \int_{S^{(b)}(\xi)} \V{n}^{(b)} \cdot (\stress^{(b)} \cdot \V{n}^{(b)})  \, dS \\
M^{(b)}(\xi) = \displaystyle - \int_{S^{(b)}(\xi)} \V{n}^{(b)} \cdot (\stress^{(b)} \cdot \V{n}^{(b)})  \,  \eta  \, dS
\end{pmatrix} \, ,
\end{equation}
the generalised distributed forces and moments and the generalised prescribed forces and moments are:
\begin{equation}
\begin{pmatrix}
\bar{\V{f}}_d  \\
m_d
\end{pmatrix} =
\begin{pmatrix}
\displaystyle \int_{S^{(b)}(\xi)} \V{f}_d  \, dS \\
\displaystyle - \int_{S^{(b)}(\xi)} \V{f}_d \cdot \V{n}^{(b)}  \eta \, dS  
\end{pmatrix} \qquad
\begin{pmatrix}
\bar{\V{F}}_d  \\
M_d
\end{pmatrix} =
\begin{pmatrix}
\displaystyle \int_{S^{(b)}(\xi)} \V{F}_d  \, d\eta \, d z \\
\displaystyle - \int_{S^{(b)}(\xi)} \V{F}_d \cdot \V{n}^{(b)}  \eta \, d\eta \, d z  
\end{pmatrix} \, ,
\end{equation}
and the generalised reaction forces and moments read:
\begin{equation}
\begin{pmatrix}
\bar{\V{F}}^{(b)}  \\
M^{(b)}
\end{pmatrix}_{|\V{P}^{(b),i}} =
\begin{pmatrix}
\displaystyle \int_{S^{(b),i} \in \mathcal{J}} \stress^{(b)} \cdot \V{n}^{(b)}   \, dS \\
\displaystyle - \int_{S^{(b),i} \in \mathcal{J}} \V{n}^{(b)} \cdot ( \stress^{(b)} \cdot \V{\tilde{n}}^{(b)} ) \,  \eta \, dS
\end{pmatrix} \, .
\end{equation}

\subsubsection{Rigid joint relations}

Let us consider a joint $(j)$ of centroid $P_i \in \mathcal{J}$ between a set of beams $\mathcal{C}^{(j)} = \{(b), \, (b'), \, (b''), \, ... \}$. We assume that the joint behaves like a rigid body. Hence the displacement $\disp^{(j)}$ belongs to space:
\begin{equation}
\mathcal{U}^{{(j)}}=\left\{\disp \in H^1(\Omega^{(j)}) \ | \  \disp_{| \V{M}} = \V{q}^{(j)} + \V{OM} \wedge \V{z} \, \theta^{(j)} \right\} \, ,
\end{equation}
Writing that the displacement field must be continuous between the joint and the connected beams, we obtain the lattice approximation of the compatibility conditions at point $P_i$:
\begin{equation}
\left\{ \begin{array}{l}
\theta^{(j)} = \theta^{(b)} \\
\V{q}^{(j)} = \V{q}^{(b)}
\end{array} \right. \quad \forall \, (b) \in \mathcal{C}^{(j)} \, .
\end{equation}

As it has been done in the previous subsection for a beam $(b)$, the equilibrium of a joint (j) reads:
\begin{equation}
\label{eq:equi_joint}
\begin{array}{l}
\displaystyle \forall \, \disptest \in \mathcal{U}^{{(j)}} \\
\displaystyle \sum_{b \in \mathcal{C}} \int_{\partial \Omega^{(j)} \cap \partial \Omega^{(b)}} \stress^{(b)} \cdot \left(-\V{\tilde{n}}^{(b)}\right) \cdot \disptest \, d \Gamma= 0 
\end{array} \, ,
\end{equation}
where the expression of the internal virtual work has vanished due to the  assumption of rigid body motion, and the volume forces are ignored because of the slenderness assumption. Therefore, only the terms associated to reaction forces remain. Integrating each of the terms in \eqref{eq:equi_joint} along the contact planes between the joint and the connected beams, we obtain the following static lattice equilibrium relations:
\begin{equation}
\left\{ \begin{array}{l}
\displaystyle \sum_{b \in \mathcal{C}^{(j)}}  (\bar{\V{F}}^{(b)} \cdot \V{q}^{(b)\star})_{| \V{P}^{(b),i} = \V{P}_j}  = 0 \\
\displaystyle \sum_{b \in \mathcal{C}^{(j)}} (M^{(b)} \, \theta^{(b)\star})_{| \V{P}^{(b),i} = \V{P}_j}  = 0
\end{array} \right.  \, ,
\end{equation}

\subsubsection{Assembled lattice problem}

\begin{figure*}[htb]
       \centering
       \includegraphics[width=0.7 \linewidth]{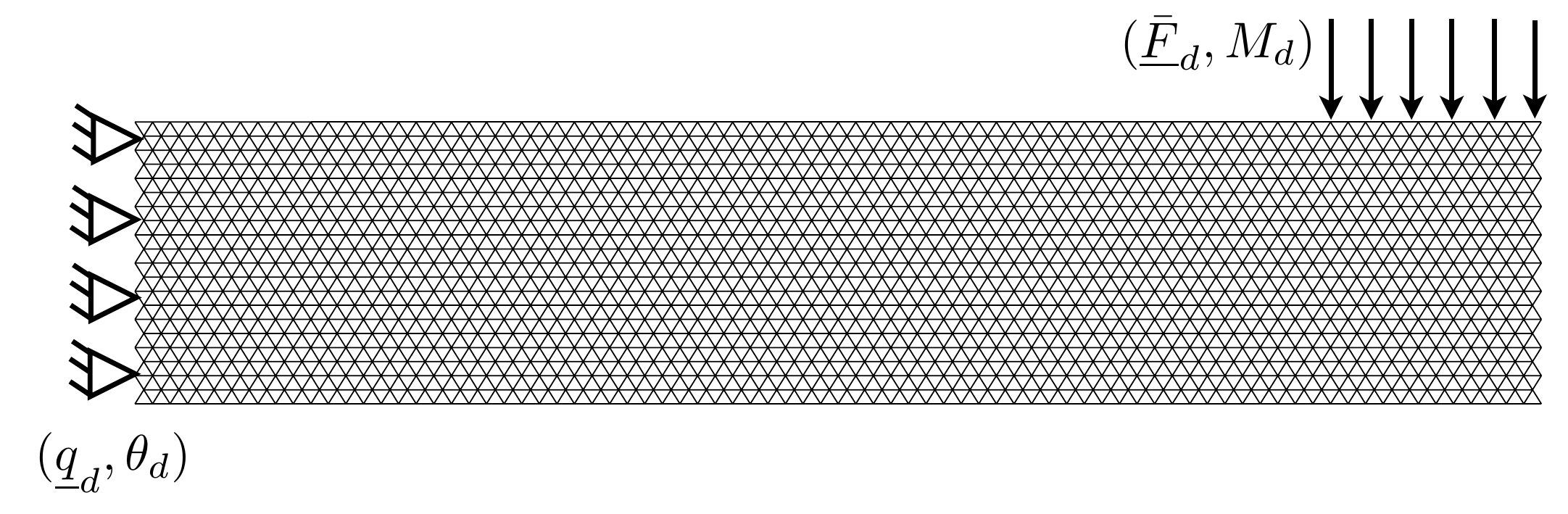}
       \caption{Periodic lattice structure with hexagonal repeated pattern as defined in figure. Example of boundary conditions are also represented.}
       \label{fig:LatticeProblem}
\end{figure*}

Summing the local weak formulation of the balance equation of the beams and taking the rigid joint relations into account, we obtain the following weak form for the lattice problem:
\begin{equation}
\label{eq:EquilibriumLatice}
\begin{array}{l}
\displaystyle \text{Find} \ S= (\V{q},\theta) \in \mathcal{Q} \ \text{such that:} \  \forall \, S^\star \in \mathcal{Q}^{0}, 
\\ \displaystyle
\sum_{b=1}^{n_b} \left( \int_{\omega^{(b)}} \bar{\V{\sigma}}^{(b)} \cdot  \bar{\V{\epsilon}}^{(b)\star} \, d \xi \right) 
= \int_{\omega} \bar{\V{f}}_d . \V{q}^\star \, d \xi  
+ \int_{\omega} m_d \, \theta^\star \, d \xi
+ \sum_{P^{(b),i} \in  \mathcal{P}_f} (\bar{\V{F}}_d . \V{q}^\star)_{| P^{(b),i}} 
\\  \displaystyle \qquad
+ \sum_{P^{(b),i} \in  \mathcal{P}_f} (M_d . \theta^\star)_{| P^{(b),i}} 
 \, ,
\end{array}
\end{equation}
where 
\begin{equation}
\displaystyle
\mathcal{Q} =  \displaystyle \left\{ (\V{q},\theta) = \bigcup_{b \in \llbracket 1,n_b \rrbracket} (\V{q}^{(b)},\theta^{(b)}) \in \mathcal{C}^{0}(\omega) 
\, | \,\forall \, b \in \llbracket 1,n_b \rrbracket , \, (\V{q}^{(b)},\theta^{(b)}) \in \mathcal{Q}^{(b)}  \right\} \, ,
\end{equation}
and $\mathcal{Q}^{0}$ is the associated vector space.

\subsection{Elastic-damageable constitutive law}

We first describe the case of linear elastic beams, establish the link between the theory of linear beam networks and continuous elastic bodies. We then extend these concepts to damageable media.

\subsubsection{Elastic properties}

Consider the linear elastic constitutive law at a point $\V{M}$ of beam $\Omega^{(b)}$:
\begin{equation}
\M{\epsilon} (\disp^{(b)}) = \frac{1+\nu}{E} \stress^{(b)} - \frac{\nu}{E} \textrm{Tr}({\stress^{(b)}}) \M{I}_d \, ,
\end{equation}
where $E$ is the Young modulus of the material, and $\nu$ is its Poisson ratio.
We assume the following energy equivalence in a section of the beam: the virtual work of the stress field in a displacement field constrained by the Euler-Bernoulli assumption is equal to the virtual work of the beam formulation, which reads
\begin{equation}
\displaystyle \forall \, \disp^{(b) \star} \in \mathcal{U}^{l,(b)}, \quad
\displaystyle \int_{S^{(b)}(\xi(\V{M}))} \stress^{(b)} : \M{\epsilon} (\disp^{(b) \star}) \, dS = \bar{\V{\sigma}}^{(b)} \cdot \bar{\V{\epsilon}}^{(b)}(S^{(b)\star}(\xi(\V{M})) \, .
\end{equation}
We obtain the classical linear state law:
\begin{equation}
\label{eq:linestate}
\begin{pmatrix}
N^{(b)} \\
M^{(b)}
\end{pmatrix} =
\begin{pmatrix}
E^{(b)} S^{(b)} & 0 \\
0 & E^{(b)} I^{(b)}
\end{pmatrix}
\begin{pmatrix}
v^{(b)}_{,\xi} \\
\theta^{(b)}_{,\xi}
\end{pmatrix} \, ,
\end{equation}
where $I^{(b)} = \int_{S^{(b)}(\xi(\V{M}))} \eta^2 \, d S$ is the second moment of area of section $S^{(b)}(\xi(\V{M}))$.

\subsubsection{Link with an elastic continuum in the case of hexagonal regular lattice}

\begin{figure}[htb]
       \centering
       \includegraphics[width=0.5 \linewidth]{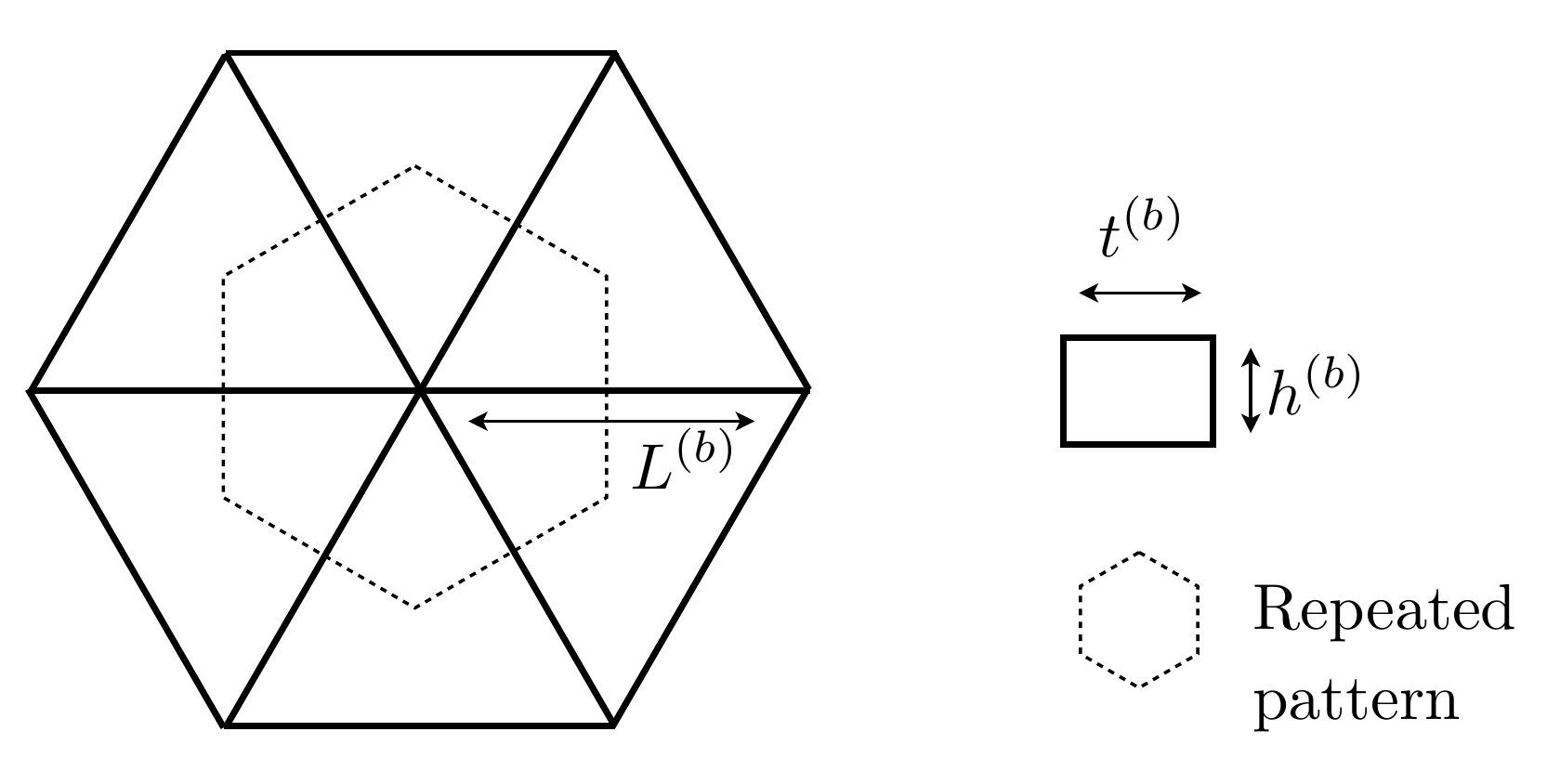}
       \caption{Periodic lattice structure with hexagonal repeated pattern}
       \label{fig:LatticeRegular}
\end{figure}

We now consider a periodic plane lattice structure made of a repeated hexagonal pattern as shown in figures \ref{fig:LatticeProblem} and \ref{fig:LatticeRegular}. The length of the beams in the network $L^{(b)}$, their thickness $h^{(b)}$ and depth $t^{(b)}$ are supposed constant trough the network. By classical homogenisation of the behaviour of the hexagonal representative volume element with respect to a point of a 2D continuum assumed in plane stress, and under the assumption of zero distributed forces, it is shown that the lattice structure is energetically equivalent to a continuum with the following material properties:
\begin{equation}
E = \left( 2 \sqrt{3} \, \frac{h^{(b)}}{ L^{(b)}} \,  \frac{\displaystyle 1+\left(\frac{h^{(b)}}{L^{(b)}}\right)^2}{\displaystyle 3+\left(\frac{h^{(b)}}{L^{(b)}}\right)^2} \, \frac{t^{(b)}}{t} \right) E^{(b)}
\, , \quad
\nu = \frac{\displaystyle 1-\left(\frac{h^{(b)}}{L^{(b)}}\right)^2}{\displaystyle 3+\left(\frac{h^{(b)}}{L^{(b)}}\right)^2} \, ,
\end{equation}
where $t$ is the depth of the continuous structure, which is not necessarily equal to the depth of the beams $t^{(b)}$.

Unfortunately, this theory, as far as the authors know, has not been extended to damage mechanics. It suffers the limitation of classical homogenisation to non-softening behaviours. In the literature, the elastic constants are used as a starting point for a phenomenological damage model, whose parameters then need to be fitted to experimental results. We follow the same approach. The damage model used in our simulations is described in the next subsection.

\subsubsection{Elastic-damage law}

The model presented here is based on classical damage mechanics \cite{chaboche1988}, applied to lattice structures. We introduce two different damage mechanisms, one acting in traction and the other one acting in bending. This assumption is consistent with the model used in \cite{karihalooshao2003}, where it is argued that damage in traction corresponds to damage due to hydrostatic deformations, while damage due to bending corresponds to shear damage.

We postulate the following Helmholtz free energy per unit length of beam $(b)$:
\begin{equation}
 \displaystyle \psi \left( \bar{\V{\epsilon}},\V{d}  \right) =  \displaystyle \frac{1}{2} \left(   E^{(b)} S^{(b)} (1-d_n) \frac{<v_{,\xi}>_+}{v_{,\xi}} \, {v_{,\xi}}^2 
\right.  \displaystyle \left. 
+ E^{(b)} I^{(b)}  (1-d_t) {\theta_{,\xi}}^2 \right) \, .
\end{equation}
$d_n$ and $d_t$ are two damage variables ranging from 0 to 1. They account for the non-reversible softening of the beam with increasing load, in respectively tension and bending. Compression in this model does not dissipate energy, which is mathematically introduced by making use of the positive part extractor $< \, . \, >_+$. We introduce the compact notation $\V{d} = (d_n \ d_t)^T$. 

The relationship between generalised stress and strain in beam $(b)$ is obtained as follows:
\begin{equation}
\begin{pmatrix}
N \\
M
\end{pmatrix} 
= \frac{\partial \psi}{\partial \bar{\V{\epsilon}}} =
\begin{pmatrix}
 \displaystyle E^{(b)} S^{(b)} (1-d_n) \frac{<v_{,\xi}>_+}{v_{,\xi}} & 0\\
 0 & E^{(b)} I^{(b)}  (1-d_t)
\end{pmatrix} 
\begin{pmatrix}
u_{,\xi}  \\
\theta_{,\xi}
\end{pmatrix}  \, .
\end{equation}
This state law is the nonlinear counterpart of the linear state law \eqref{eq:linestate}.

The second state law, which links the damage variables to dual driving thermodynamic forces, reads:
\begin{equation}
\begin{pmatrix}
Y_n \\
Y_t
\end{pmatrix}
= - \frac{\partial \psi}{\partial \V{d}} = 
\frac{1}{2} 
\begin{pmatrix}
\displaystyle E^{(b)} S^{(b)} \frac{<v_{,\xi}>_+}{v_{,\xi}} \, {v_{,\xi}}^2 \\
\displaystyle E^{(b)} I^{(b)}  {\theta_{,\xi}}^2
\end{pmatrix} \, .
\end{equation}

At last, an evolution law is defined to fully define the damage evolution:
\begin{equation}
Y(t) = \max_{\tau \leq t} \left( \left( Y_n (\tau) \right)^\alpha + \left( \gamma Y_t (\tau) \right)^\alpha \right)^{\frac{1}{\alpha}} \, ,
\end{equation}
\begin{equation}
d_n = d_t = \min \left(  \left( \frac{n}{n+1} \, \frac{ < Y-Y_0 >_+}{Y_c -Y_0} \right)^{n}, \, 1 \right) \, .
\end{equation}

This damage law is inspired by the model described in \cite{allixleveque1998} for composite laminates. We refer to this work for a comprehensive interpretation of the different parameters. We will just notice that $Y$ is an equivalent damage energy release rate which governs the evolution of damage with traction and banding. The critical value $Y_c$ is therefore the ``strength'' of the beam section.

\subsection{Randomly distributed material properties}

The damageable lattice model is used to derive a three-phase model for concrete. Such models consider three different entities: matrix (cement), inclusions (hard particles, assumed spherical) and an interface between these two entities (see \cite{scrivenercrumbie2004} for an evidence of the existence of such interface). Plane $(O,\V{x},\V{y})$ is a section of the three dimensional particulate composite structure. A projection of the material properties onto the lattice model is performed as follows:
\begin{itemize}
\item The size and position of the spherical particles is generated randomly, using the algorithm given below
\item Beams whose two extremities belong to either the matrix phase or a single inclusion will be attributed respectively the properties of the matrix or of the particles
\item A beam which has its two extremities in different phases will have material properties corresponding to the interface.
\end{itemize}

The particle distribution is generated in a similar way to the one used in \cite{vanmiervanvliet2002,karihalooshao2003,grasslbazant2009}. The probability of an arbitrary point of the plane $(O,\V{x},\V{y})$, located in any of the inclusions of the structure, to belong to an inclusion whose section by the plane has a diameter smaller than $D$ is given the cumulative distribution function $P_c(D)$ identified in \cite{walraven1980}. If we define the normalised diameter by $\bar{D}= \frac{D}{D_{\textrm{Max}}}$, where $D_{\textrm{Max}}$ is the maximum particle diameter, this distribution function $P_c(D)$ is given by the following equation:
\begin{equation}
\label{eq:distri}
\begin{array}{cl}
\displaystyle 
P_c(D) = & \displaystyle 
1.065 \, \left( \frac{D}{D_{\textrm{Max}}} \right)^{0.5}
-
0.053 \, \left( \frac{D}{D_{\textrm{Max}}} \right)^{4} 
\\ & \displaystyle
-
0.012 \, \left( \frac{D}{D_{\textrm{Max}}} \right)^{6} 
-
0.0045 \, \left( \frac{D}{D_{\textrm{Max}}} \right)^{8}
-
0.0025 \, \left( \frac{D}{D_{\textrm{Max}}} \right)^{10}
\end{array}
\end{equation}
If we look for a distribution of particles with a discrete set of projected diameters $\mathcal{D} = \{D_i\}_{i \in \llbracket 1,n_D\rrbracket}$ ordered in decreasing order such that $D_i - D_{i+1} = \Delta D$ (with $\Delta D = \frac{D_\textrm{Max}}{n_D}$ a positive number), we can obtain that the number $n_i$ of particles of projected diameter $D_i$ is given by:
\begin{equation}
\forall i \in \llbracket 1,n_D\rrbracket , \quad   n_i = \frac{A \, P_k}{\pi \left( \frac{D_i}{2} \right)^2} \left( P_c \left( \frac{D_i+\frac{\Delta D}{2}}{D_{\textrm{Max}}} \right) - P_c \left( \frac{D_i-\frac{\Delta D}{2}}{D_{\textrm{Max}}} \right) \right) \, 
\end{equation}
where $A$ is the surface of the structure and $P_k$ is the volume fraction of inclusions in the particulate composite. 

To position the particles, we randomly draw the coordinates of the projected centre of each one of them, in turn, starting from the particles of largest projected diameter. To avoid particle collision, we enforce the constraint that the distance between the projected centre of any two particles must at least equal to $1.1 \, \left( \frac{D_i}{2} + \frac{D_j}{2} \right)$, where $D_i$ and $D_j$ are the projected diameters of the two arbitrary particles. This constraint is enforced numerically by simply drawing an other random position if the currently positioned particle does not satisfy this condition.
 
 \begin{figure}[htb]
        \centering
        \includegraphics[width=1.0 \linewidth]{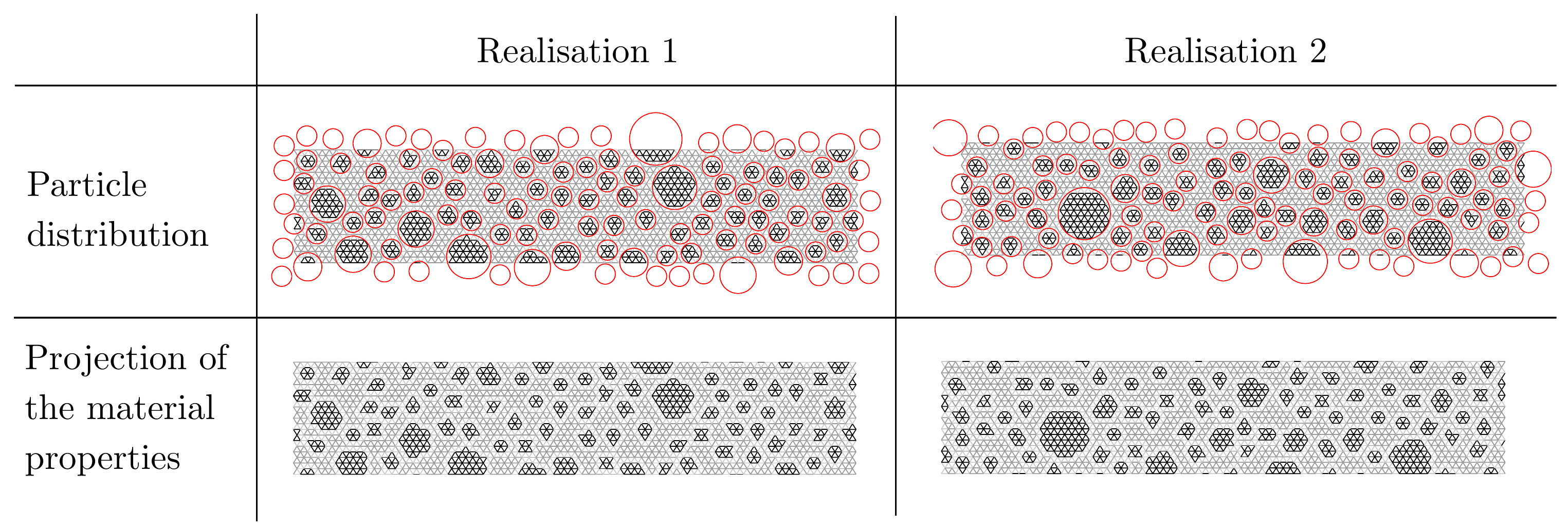}
        \caption{Two realisations of the random distribution of the material properties. The inclusions are spherical, and their projection onto the lattice structure is represented in black. The grey bars correspond to the matrix of the particulate composite, while the light grey ones define the weak interface between aggregates and matrix.}
        \label{fig:ParticleDistribution}
 \end{figure}
 
  The material properties of the three phases are attributed as follows:
  \begin{itemize}
  \item a beam that belongs to an inclusion cannot be damaged ($Y_{c,\textrm{Inc}} = \infty$ in the damage model). Its elasticity constants are $E_\textrm{Inc} $ and $\nu_\textrm{Inc}$.
  \item a beam that belongs to the matrix phase has elasticity constants $E_\textrm{Mat}$ and $\nu_\textrm{Mat}$, and damage parameters $\alpha_\textrm{Mat}$, $\gamma_\textrm{Mat}$, $Y_{c,\textrm{Mat}} $, $Y_{0,\textrm{Mat}}$ and $n_\textrm{Mat}$.
    \item a beam that belongs to the interface phase has elasticity constants $E_\textrm{Inter}$ and $\nu_\textrm{Inter}$, and damage parameters $\alpha_\textrm{Inter}$, $\gamma_\textrm{Inter}$, $Y_{c,\textrm{Inter}} $, $Y_{0,\textrm{Inter}}$ and $n_\textrm{Inter}$. The interface is, in general terms, softer and weaker than the matrix.
  \end{itemize}
 
The main drawback of such a model is that it is based on the assumption of plane stress, which is of course hardly justified in the case of spherical inclusions. However, a number of numerical investigations \cite{vanmiervanvliet2002,karihalooshao2003} have shown that this type of model is capable of giving a relatively accurate description of fracture in concrete structures. In the present study, we are more interested in the numerical behaviour of the damage model rather than in his predictive ability. We believe that the results of this paper can be extended to other damage models, whether related to concrete structures or not, as long as it exhibits the behaviour that will be described later on (e.g.: localisation of the damage process).
 
In order to describe the randomness of the model formally, we introduce the probability space $\mathcal{H} = (\Theta,\mathcal{F},P)$ for the random distribution of the material properties of the lattice network. $\Theta$ is the ensemble of outcomes of the random generation, $\mathcal{F}$ is the corresponding Sigma-algebra of subsets of $\Theta$ and $P$ is the associated probability measure. Any random distribution is characterised by distribution function \eqref{eq:distri}, a fixed maximum inclusion diameter $D_\textrm{Max}$, a fixed phase ratio, fixed material properties for each of the phases, and a given lattice network.

Two realisations of the random distribution of material properties are illustrated in figure \ref{fig:ParticleDistribution}.

\subsection{Discretisation}

We now present briefly the discretisation technique of the lattice balance equations \eqref{eq:EquilibriumLatice}. The displacement of a point of the neutral fibre of beam $(b)$ is approximated using a unique finite element, of first order degree for the normal displacement, and third order degree for the deflection:
\begin{equation} 
\begin{pmatrix}
v^{(b)} \\
w^{(b)}
\end{pmatrix} (\xi)
= 
\begin{pmatrix}
\M{\Phi}_v^{(b)}(\xi) \\
\M{\Phi}_w^{(b)}(\xi)
\end{pmatrix} 
\begin{pmatrix}
v^{(b)}(0) \\
w^{(b)}(0) \\
\theta^{(b)}(0) \\
v^{(b)}(L^{(b)}) \\
w^{(b)}(L^{(b)}) \\
\theta^{(b)}(L^{(b)})
\end{pmatrix} 
\end{equation} 
\begin{equation}
\nonumber 
\begin{pmatrix}
v^{(b)} \\
w^{(b)}
\end{pmatrix} (\xi) 
= \M{\Phi}^{(b)} (\xi) \, \V{\mathbf{V}}^{(b)}  \, ,
\end{equation}
where the matrix of shape functions is given by:
\begin{equation}
\M{\Phi} (\xi) = \begin{pmatrix}
\Phi_{v,1}(\xi) & 0 \\
0 & \Phi_{w,1}(\xi) \\
0 & \Phi_{w,2}(\xi) \\
\Phi_{v,2}(\xi) & 0 \\
0 & \Phi_{w,3}(\xi) \\
0 & \Phi_{w,4}(\xi)
\end{pmatrix}^T
\qquad  \textrm{with} \qquad
\left\{
\begin{array}{l}
\displaystyle \Phi_{v,1}(\xi) = 1 - \frac{1}{L} \xi \\
\displaystyle \Phi_{v,2}(\xi) = \xi \\
\displaystyle \Phi_{w,1}(\xi) =  \frac{2}{L^3} \xi^3- \frac{3}{L^2} \xi^2+ 1 \\
\displaystyle \Phi_{w,2}(\xi)  = \frac{1}{L^2}  \xi^3 - \frac{2}{L} \xi^2 + \xi     \\
\displaystyle \Phi_{w,3}(\xi)  = - \frac{2}{L^3} \xi^3 + \frac{3}{L^2} \xi^2  \\
\displaystyle \Phi_{w,4}(\xi)  = \frac{1}{L^2} \xi^3 - \frac{1}{L} \xi^2
\end{array} \right. \, .
\end{equation}

Hence, the local finite element approximation of the solution fields in bar $b$ can be expressed in the global basis $\mathcal{B}$ using the transformation:
\begin{equation} 
\begin{pmatrix}
\displaystyle q_x^{(b)} \\
\displaystyle q_y^{(b)} \\
\displaystyle \theta^{(b)}
\end{pmatrix} (\xi)
= 
\begin{pmatrix}
\displaystyle {\M{R}^{(b)}}^T \M{\Phi}^{(b)} (\xi) \\
\displaystyle \M{\Phi}_{w,\xi}^{(b)}(\xi)
\end{pmatrix}
\V{\mathbf{V}}^{(b)} \, ,
\end{equation}
where $\V{q}$ is decomposed in $\mathcal{B}$ in the form $\V{q} = q_x \, \V{x} + q_y \, \V{y} $ and $\M{R}^{(b)}$ is the following rotation matrix:
\begin{equation} 
\M{R}^{(b)} = 
\begin{pmatrix}
({\V{n}^{(b)}})^T  \V{x} & ({\V{n}^{(b)}})^T  \V{y} \\
-({\V{n}^{(b)}})^T \, \V{y} & ({\V{n}^{(b)}})^T \V{x}
\end{pmatrix} \, .
\end{equation}
The approximated generalised strain used above reads:
\begin{equation} 
\begin{pmatrix}
v^{(b)}_{,\xi} \\
\theta^{(b)}_{,\xi}
\end{pmatrix} (\xi)
= 
\begin{pmatrix}
\M{\Phi}_{v,\xi}^{(b)}(\xi) \\
\M{\Phi}_{w,\xi \xi}^{(b)}(\xi)
\end{pmatrix}
\V{\mathbf{V}}^{(b)} \, .
\end{equation}

In order to ensure the $\mathcal{C}^{0}$ continuity of the solution fields $(\V{q},\theta)$ over $\omega$, we introduce a unique vector of nodal unknowns $ \dispdiscr = \begin{pmatrix} {q_x}_{|P_1} & {q_y}_{|P_1} &  {\theta}_{|P_1} & ... & {q_x}_{|P_{n_p}} & {q_y}_{|P_{n_p}} &  {\theta}_{|P_{n_p}} \end{pmatrix}^T$, with $\dispdiscr \in \mathbb{R}^{n_u}$. The vector of nodal unknowns of beam $b$ is denoted by $\dispdiscr^{(b)}$ and is obtained from $\dispdiscr$ by the following extraction:
\begin{equation}
\dispdiscr^{(b)} = \M{\mathbf{A}}^{(b)} \dispdiscr
\end{equation}
The relationship between $\V{\mathbf{V}}^{(b)}$ and $\V{\mathbf{U}}^{(b)}$ is:
\begin{equation}
\V{\mathbf{V}}^{(b)} = \M{\mathbf{R}}^{(b)} \V{\mathbf{U}}^{(b)}
\end{equation}
\begin{equation}
\nonumber
\M{\mathbf{R}}^{(b)} = 
\begin{pmatrix}
\M{R}^{(b)} & \M{0} & \M{0} & \M{0} \\
\M{0} & 1 & \M{0} & 0 \\
\M{0} & \M{0} & \M{0} & \M{R}^{(b)} \\
\M{0} & 0 & \M{0} & 1
\end{pmatrix} \, .
\end{equation}

We finally obtain the expression of the local solution fields and strains as a function of the vector of nodal unknowns:
\begin{equation}
\begin{pmatrix}
\displaystyle q_x^{(b)} \\
\displaystyle q_y^{(b)} \\
\displaystyle \theta^{(b)}
\end{pmatrix} 
=  
\begin{pmatrix}
\displaystyle {\M{R}^{(b)}}^T \M{\Phi}^{(b)} \\
\displaystyle \M{\Phi}_{w,\xi}^{(b)}
\end{pmatrix}
\M{\mathbf{R}}^{(b)} \M{\mathbf{A}}^{(b)} \dispdiscr = \M{N}^{(b)} \M{\mathbf{A}}^{(b)} \dispdiscr
\end{equation}
\begin{equation}
\begin{pmatrix}
v^{(b)}_{,\xi} \\
\theta^{(b)}_{,\xi}
\end{pmatrix} (\xi) 
= 
\begin{pmatrix}
\M{\Phi}_{v,\xi}^{(b)} \\
\M{\Phi}_{w,\xi \xi}^{(b)}
\end{pmatrix}
\M{\mathbf{R}}^{(b)} \M{\mathbf{A}}^{(b)} \dispdiscr = \M{B}^{(b)} \M{\mathbf{A}}^{(b)} \dispdiscr
\end{equation}

By substitution of the finite element approximation into the balance equations \eqref{eq:EquilibriumLatice} for solution $S$ and test vector $S^\star$ (Galerkin framework), we obtain the semi-discrete system of $n_u$ time-dependent equations:
\begin{equation}
\displaystyle \forall \, t \in \mathcal{T}, \,  \text{find} \, \dispdiscr \in \mathbb{R}^{n_u} \ \text{such that: }  \quad
\left\{ \begin{array}{l}
\begin{array}{l}
\displaystyle \forall \, \dispdiscr^\star \in \mathbb{R}^{n_u} \ \text{ such that } \dispdiscr^\star(\mathcal{P}_u) = \zerodiscr,  \\
 \qquad \left(\dispdiscr^\star\right)^T \left( \fint \left( ( \dispdiscr_{| \tau} )_{\tau \in [0,t]}, \, \theta \right) +  \fext(t) \right) = 0 \, , \quad 
 \end{array}
\\
\ \, \displaystyle \dispdiscr(\mathcal{P}_u) = \dispdiscr_d \, ,
\end{array} \right.
\end{equation}
where the vector of internal forces, which depends on the particle distribution, is given by:
\begin{equation}
\fint \left( \left(\dispdiscr_{| \tau} , \, \theta \right)_{\tau \in [0,t]} \right) 
= - 
\sum_{b=1}^{n_b} {\M{\mathbf{A}}^{(b)}}^T \int_{\omega^{(b)}} {\mathcal{B}^{(b)}}^{T} \cdot \bar{\V{\sigma}}^{(b)}    \, d \xi
\, ,
\end{equation}
and the vector of external forces is
\begin{equation}
\displaystyle
\fext(t) 
=  \sum_{b=1}^{n_b} {\M{\mathbf{A}}^{(b)}}^T  \int_{\omega^{(b)}}  {\mathcal{N}^{(b)}}^{T} \cdot 
\begin{pmatrix}
\bar{\V{f}}_d \cdot \V{x} \\
\bar{\V{f}}_d \cdot \V{y} \\
m_d
\end{pmatrix}
\, d \xi 
+ \sum_{b=1}^{n_b} \sum_{P^{(b),i} \in  \mathcal{P}_f} {\mathcal{N}^{(b)}}_{|P^{(b),i}}^{T} \cdot 
\begin{pmatrix}
\bar{\V{F}}_d \cdot \V{x} \\
\bar{\V{F}}_d \cdot \V{y} \\
M_d
\end{pmatrix}
\, .
\end{equation}
The two integral terms are evaluated for each beam using a 3-point Gauss quadrature rule.

%% file: solution.tex
For a given distribution of the material properties, the nonlinear solution strategy used to solve the problem over time is a classical time discretisation scheme for quasi-static, rate-independent problems, associated with the continuation algorithm proposed by \cite{lorentzbadel2004} to handle the instabilities, which are a classical feature of fracture mechanics. This procedure consists in integrating the history of the irreversible process by looking for a set of consecutive solutions at some pseudo-times in a discrete time space $\mathcal{T}^h = (t_n)_{n \in \llbracket 0 , n_t \rrbracket}$ (with $t_0=0$ and $t_{n_t}=T$), together with the unknown amplitude of the external load.

The introduction of the time discretisation into equation \eqref{eq:equilibrium} at any time $t_n \in \mathcal{T}^h$ of the analysis, and for a given distribution $\theta$ of the material properties, leads to the following fully discrete nonlinear vectorial equation:
\begin{equation}
\label{eq:discretized_equilibrium}
\fint \left( \left( \dispdiscr_{|t} \right)_{t \in \{ t_1, \, ... \, t_n \} } , \theta \right) + \lambda(t) \, \frac{\fext(t)}{\| \fext(t) \|_2}= \zerodiscr
\end{equation}
Homogeneous Dirichlet boundary conditions have been assumed at this stage for the sake of concision.

In order to close the system of equations and define the time evolution of $\lambda$, an additional constraint is introduced. It enforces that the maximum dissipation over all quadrature points of the lattice structure is equal to a prescribed value. The system comprising the balance equations and the additional constraints are solved by a Newton algorithm with line search.







 \subsection{Motivation of the work by a test problem}
 
\begin{figure}[p]
         \centering
         \includegraphics[width=1.0 \linewidth]{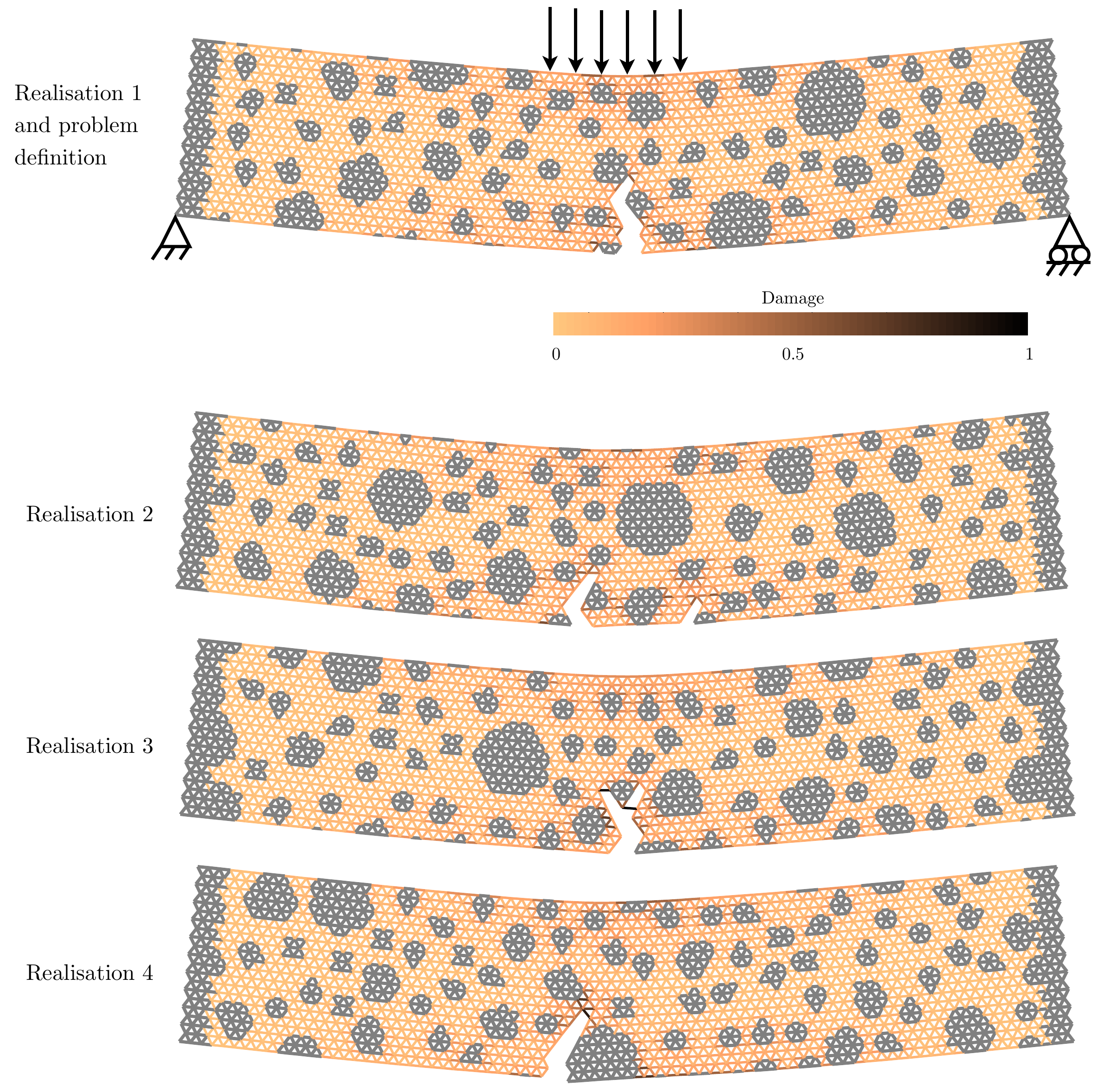}
         \caption{Definition of the test problem of fracture in random particulate composites (top), and solutions corresponding to different realisations of this problem}
         \label{fig:RandomCracks}
  \end{figure}

 An example for the problem of fracture of random composites is depicted in figure \ref{fig:RandomCracks}. The structure considered is a 2D damageable heterogeneous rectangular lattice beam. Homogeneous Dirichlet boundary conditions are applied at the bottom left and right-hand corners of the structure. Compressive vertical forces are applied on the top part of its boundary, as illustrated in the figure (\ref{fig:RandomCracks}, top). A crack initiates in the region where the generalised stress is maximum, and propagates, broadly in the vertical direction, to minimise the potential energy stored by the structure. The time history of the irreversible damage mechanisms is integrated using 50 time steps. The picture shows the solution obtained at the last time step of the analysis. The aggregates are represented in grey, while the colorscale in the matrix and interface indicate the level of damage in each of the beams comprising the lattice.
 
We have used in our test cases the values $Y_{c,\textrm{Inter}} = \frac{1}{4} Y_{c,\textrm{Mat}}$, which is why the crack tends to follow the  aggregate boundaries. The elastic constants of the three phases are all equal. Therefore, the heterogeneity of the problem is only due to damage, and is therefore relatively localised. This restriction, which is not physical as the aggregates are stiffer than the matrix in real concrete structures, will be justified later on.
 
 The solution corresponding to three other particle distributions are presented below. One can see the crack paths differ, but that damage localises in a relatively small region compared to the size of the beam. Away from this so-called ``process zone'', the solution of the four numerical experiments seem qualitatively close, which justifies the search of invariant properties for the random process.
 
The idea is then to look for an approximate of solution $(\dispdiscr(t,\theta))$ at any time $t \in \mathcal{T}^h$ and for any realisation $\theta \in \Theta$ of the random distribution of material properties in a deterministic spatial subspace of small dimension $n_\phi$, spanned by global basis vectors $\left( (\phi_i)_{i  \in \llbracket 1, n_\phi \rrbracket} \right) \in (\mathbb{R}^{n_u})^{n_\phi}$. This so-called reduced space being determined, one could possibly construct a reduced order model for the fast solution of the problem with randomness. The idea is to look for $(\dispdiscr(t,\theta))$ in the space spanned by the pre-determined global basis vectors. In doing so, the numerous spatial unknowns of the discrete lattice problem can be reduced to $n_\phi$ amplitudes associated with the global shape functions $\left( (\phi_i)_{i \in \llbracket 1, n_\phi \rrbracket} \right)$, which potentially allows for orders of magnitude of gain in numerical efficiency.

The scope of this article is not the construction of the reduced order model itself, for which we refer to the extensive literature concerning reduced order modelling in the linear and mildly nonlinear case (see the references in the introduction of this article). In the case of heterogeneous structures, and by extension in the case of nonlinear heterogeneous structures, the extraction of the relevant reduced space itself is not established. This first step towards the construction of a reduced model for fracture in random materials is the topic of the paper.

%% file: POD.tex
\subsection{Proper orthogonal decomposition (POD)}

The proper orthogonal decomposition (POD) \cite{pearson1901,hotelling1933,abdiwilliams2010} is a particular family of transforms that aim at extracting deterministic trends from randomly scattered data. Such transforms are powerful tools for the analysis of parametric problems and problems with randomness are used in numerous field of applications (see for instance \cite{sirovich1987} and the review and analysis of the different variants of the POD proposed in \cite{lianglee2002}).  In the context of the analysis of multivariate random processes (or time series), the POD proposes to approximate the random process of interest as a combination of simply structured random processes. In order to do so, the random process is expanded as a finite sum of orthogonal deterministic spatial vectors weighted by scalar random processes. As opposed to classical Fourier transforms, the deterministic functions are not defined \textit{a priori}, but found \textit{a posteriori} by solving an optimisation problem. 

We look for such an approximation of the stochastic time evolution of the nodal values of the displacement field in the damageable lattice structure. The POD expansion of $\dispdiscr$ at order ${n_\phi}$ is a random process in the form
\begin{equation}
\label{eq:SepVariables}
\begin{array}{rcl}
\displaystyle \bar{\dispdiscr}  : \quad \mathcal{T}^h \times \Theta & \displaystyle \rightarrow & \mathbb{R}^{n_u}
\\ \displaystyle
\displaystyle (t,\theta) & \displaystyle \mapsto & \displaystyle \sum_{i=1}^{n_\phi} \V{\boldsymbol{\phi}}_i \, \alpha_i(t,\theta)  \, ,
\end{array}
\end{equation}
where for all $i \in \llbracket 1, n_\phi \rrbracket$, $\V{\boldsymbol{\phi}}_i \in \mathbb{R}^{n_u}$ is a deterministic ``space'' vector, while $\alpha_i$ is the associated scalar random process, called ``weight''.  $n_\phi \leq n_u$ is the order of the POD expansion. 

The POD proposes to look for the optimal approximation $\bar{\dispdiscr}$ that minimises a distance $d(\dispdiscr,\bar{\dispdiscr})$ between the expansion and the exact random process of interest $\dispdiscr$. 
In order to define this distance, let us introduce the space $\mathcal{S}$ of multivariate random processes of $n_u$-dimension random vectors defined on $\mathcal{H}$ and indexed in $\mathcal{T}^h$. A natural scalar product $<. \, , \, .>_{\mathcal{S}}$ on $\mathcal{S}$, and an associated norm $\| \, . \, \|_{\mathcal{S}}$, are defined as follows:
\begin{equation}
\begin{array}{l}
\displaystyle <\V{\mathbf{X}},\V{\mathbf{Y}}>_{\mathcal{S}} = E \left( \sum_{t \in \mathcal{T}^{h}} \V{\mathbf{X}}(t,\theta)^T \V{\mathbf{Y}}(t,\theta) 
\right) = \sum_{\theta \in \Theta} P(\theta) \left( \sum_{t \in \mathcal{T}^{h}} \V{\mathbf{X}}(t)^T \V{\mathbf{Y}}(t) \right) \\
\displaystyle  \| \V{\mathbf{X}} \|_{\mathcal{S}} = \left( <\V{\mathbf{X}},\V{\mathbf{X}}>_{\mathcal{S}} \right)^\frac{1}{2}
\, ,
\end{array}
\end{equation}
where $P(\theta)$ is the probability for elementary event $\theta \in \Theta$ to happen and $\V{\mathbf{X}}$ and $\V{\mathbf{Y}}$ are arbitrary elements of $\mathcal{S}$.
Let us introduce the matrix notation $\displaystyle \sum_{i=1}^{n_\phi} \V{\boldsymbol{\phi}}_i \, \alpha_i(t) = \basereduc \, \coeffreduc(t)$, for all $t \in  \mathcal{T}^h$. The columns of operator $\basereduc \in \mathbb{R}^{n_u \times n_\phi}$ are the deterministic space vectors $\left( \V{\boldsymbol{\phi}}_i \right)_{i \in \llbracket 1, n_\phi \rrbracket}$, and $\coeffreduc$ is the column vector of scalar random processes $\left( \alpha_i \right)_{i \in \llbracket 1, n_\phi \rrbracket}$. With these compact notations, and calling $\mathcal{S}^{n_\phi}$ the subspace of $\mathcal{S}$ comprising all random processes of form \eqref{eq:SepVariables}, the POD looks for an optimal decomposition \eqref{eq:SepVariables} which satisfies
\begin{equation}
\label{eq:MetricPOD}
\bar{\dispdiscr} = \underset{\bar{\dispdiscr}^\star \in \mathcal{S}^{n_\phi}}{\textrm{argmin}} \left( d_{\mathcal{S}}(\dispdiscr,\bar{\dispdiscr}^\star) \right) =  \underset{\bar{\dispdiscr}^\star \in \mathcal{S}^{n_\phi}}{\textrm{argmin}} \left( \| \dispdiscr - \bar{\dispdiscr^\star} \|_{\mathcal{S}}^2 \right) \, ,
\end{equation}
under the constraint of orthonormality of the space deterministic vectors
\begin{equation}
\label{eq:orthonorm}
\basereduc^T \basereduc = \M{\mathbf{I}}_d \, .
\end{equation}

The solution to optimisation problem (\ref{eq:MetricPOD},\ref{eq:orthonorm}) requires the introduction of the POD  operator (covariance operator if the random process is centred)
\begin{equation}
\M{\mathbf{H}} = E \left( \sum_{t \in \mathcal{T}^h} \dispdiscr(t,\theta) \, \dispdiscr(t,\theta)^T \right) \, .
\end{equation}
A global minimiser of (\ref{eq:MetricPOD},\ref{eq:orthonorm}) is then given by the following POD expansion:
\begin{itemize}
\item the scalar random processes are such that $\coeffreduc(t,\theta) =  \basereduc^T  \, \dispdiscr(t,\theta) $, for all $(t,\theta) \in  \mathcal{T}^h \times \Theta$, which implies that $\bar{\dispdiscr}(t,\theta)=\basereduc \, \basereduc^T \dispdiscr(t,\theta)$ is the orthogonal projection of ${\dispdiscr}(t,\theta)$ onto $\textrm{Im}(\basereduc)$,
\item the spacial modes $\left( \V{\boldsymbol{\phi}}_i \right)_{i \in \llbracket 1, n_\phi \rrbracket}$ are the eigenvectors of $\M{\mathbf{H}}$ associated to its largest $n_\phi$ eigenvalues $(\lambda_i)_{i \in \llbracket 1, n_\phi \rrbracket}$ ($\M{\mathbf{H}}$ has $n_u$ real and non-negative eigenvalues $(\lambda_i)_{i \in \llbracket 1, n_u \rrbracket}$ as $\M{\mathbf{H}}$ is real and symmetric).
\end{itemize}

The approximation error (i.e. the error due to the truncation of the expansion at order $n_\phi$) for a proper orthogonal decomposition $\bar{\dispdiscr}$ of $\dispdiscr$ is
\begin{equation}
\label{eq:nu}
\nu = \sqrt{ d_{\mathcal{S}} \left( \dispdiscr,\bar{\dispdiscr}  \right) } = \sqrt{ \sum_{i=n_\phi+1}^{n_u} \lambda_i } \, .
\end{equation}

For the POD to be useful in practice, the error of the decomposition should decrease quickly with order $n_\phi$. In other words, the state space of the solutions of the stochastic problem of fracture over $\mathcal{P} = \mathcal{T}^h \times \Theta$ should be well-approximated in a low-dimensional manifold $\textrm{Im}(\basereduc)$ of $\mathbb{R}^{n_u}$.

\subsection{Empirical Proper orthogonal decomposition}

In practice, the POD operator cannot be computed exactly. It is replaced by the usual statistical estimate \cite{sirovich1987}
\begin{equation}
\M{\mathbf{H}}^s = \frac{1}{n_\theta} \sum_{\theta \in \Theta^s} \left( \sum_{t \in \mathcal{T}^h} \dispdiscr(t,\theta) \, \dispdiscr(t,\theta)^T \right) \, .
\end{equation}
In the previous equation,  $\Theta^s$ is a subset of $\Theta$ containing a relatively small number $n_\theta$ of realisations of the random distribution of material properties. The set of associated realisations of the random process $\left\{ \dispdiscr(t,\mu) \right\}_{ \, (t,\mu) \in  \mathcal{T}^h \times \Theta^{s} } $ is usually called a snapshot. Let us define for simplicity $\mathcal{P}^s=\mathcal{T}^h \times \Theta^{s}$. The empirical spatial basis $\basereduc^s$ of rank $n_\phi$ obtained by extracting the eigenvectors of $\displaystyle \M{\mathbf{H}}^{s}$ associated to its largest $n_\phi$ eigenvalues $(\lambda_i^s)_{i \in \llbracket 1, n_\phi \rrbracket}$ minimises the empirical POD functional
\begin{equation}
\label{eq:snapmetric}
J^s(\basereduc^s) = \frac{1}{n_\theta} \, \frac{1}{n_t} \, \sum_{(t,\theta) \in \mathcal{P}^s} \| \dispdiscr(t,\theta) - \basereduc^s \, {\basereduc^s}^T \dispdiscr(t,\theta) \|_2^2 \, ,
\end{equation}
under the constraint of orthonormality $\eqref{eq:orthonorm}$. Notice that the spatial modes are now random, the estimate $\M{\mathbf{H}}^s$ being itself a random operator. The challenge consists in ensuring that a certain distance between the image of the empirical projector $\basereduc^s \, {\basereduc^s}^T$ and the image of the deterministic projector $\basereduc \, \basereduc^T$ is small enough for the approximation to be of some use.

Classically, the elements of the snapshot need to be normalised in some way to make the analysis easier, or to obtain results that are more consistent with the way the POD transform will be used later on. In our case, the elements of the snapshot of large amplitude tend to ``attract'' the empirical projector $\basereduc^s \,  {\basereduc^s}^T$, which is an undesirable property. We here choose to modify functional $\eqref{eq:snapmetric}$ to avoid this inconvenience:
\begin{equation}
J^{s,norm}(\basereduc^s) = \frac{1}{n_\theta} \, \frac{1}{n_t} \, \sum_{(t,\theta) \in \mathcal{P}^s} \left\| \frac{\dispdiscr(t,\theta)}{\| \dispdiscr(t,\theta) ||_2} - \basereduc^s \,  {\basereduc^s}^T \frac{\dispdiscr(t,\theta)}{\| \dispdiscr(t,\theta) ||_2} \right\|_2^2 \, .
\end{equation}
The normalised empirical POD transform obtained in this way is a weighted POD of the original, non-normalised, elements of the snapshot as shown by the equivalent form
\begin{equation}
\label{eq:snapmetricnorm}
J^{s,norm}(\basereduc^s) = \frac{1}{n_\theta} \, \frac{1}{n_t} \,  \sum_{(t,\theta) \in \mathcal{P}^s} \frac{1}{\| \dispdiscr(t,\theta) ||^2_2} \left\| \dispdiscr(t,\theta) - \bar{\dispdiscr}^{s} (t,\theta) \right\|_2^2 \, ,
\end{equation}
where $\forall \, (t,\theta) \in \mathcal{P}$, $\bar{\dispdiscr}^{s} (t,\theta) = \basereduc^s \,  {\basereduc^s}^T \, \dispdiscr(t,\theta)$ is the orthogonal projection of an arbitrary realisation of the random process onto the empirical reduced space $\textrm{Im}({\basereduc^s})$.
The empirical POD operator, modified by the proposed normalisation, reads
\begin{equation}
 \M{\mathbf{H}}^{s,norm} = \frac{1}{n_\theta} \, \frac{1}{n_t} \, \sum_{(t,\theta) \in \mathcal{P}^s} \frac{1}{\| \dispdiscr(t,\theta) ||^2_2} \, \dispdiscr(t,\theta)  \,\dispdiscr(t,\theta)^T
 \end{equation}

Let us now define the truncation error for the normalised empirical snapshot POD of order $n_\phi$
\begin{equation}
\nu^s  \displaystyle  
 = \sqrt{ \displaystyle J^{s,norm}(\basereduc^s) }
 = \sqrt{ \displaystyle\sum_{i=n_\phi+1}^{n_u} \lambda_i^s  } \, .
\end{equation}
This error estimate provides information about the distance between the realisations of the random process and their projections in the empirical reduced space $\textrm{Im}(\basereduc^s)$ for random distributions of the material properties that belong to sample space $\Theta^s$. This property and its consequences in terms of predictivity will be discussed in the next subsection.

We redefine in a similar manner the deterministic error estimate $\nu$ (equation \eqref{eq:nu}) such that it incorporates the modifications proposed in this subsection for its empirical counterpart (normalisation):
\begin{equation}
\displaystyle  \nu  \displaystyle  
 = \sqrt{J^{norm}(\basereduc)}= \sqrt{ \displaystyle\sum_{i=n_\phi+1}^{n_u} \lambda_i } \, ,
\end{equation}
where $(\lambda_i)_{i \in \llbracket 1 , n_u \rrbracket}$ are now the eigenvalues of $ \displaystyle 
\M{\mathbf{H}} = E \left( \frac{1}{n_t} \, \sum_{t \in \mathcal{T}^h}  \frac{1}{\| \dispdiscr(t,\theta) ||^2_2} \, \dispdiscr(t,\theta) \, \dispdiscr(t,\theta)^T \right) $ (in decreasing order), and the deterministic spatial basis $\basereduc$ composed of its $n_\phi$ first eigenvectors minimises 
\begin{equation}
\label{eq:snapmetric3}
J^{norm}(\basereduc) = E \left( \frac{1}{n_t} \, \sum_{t \in \mathcal{T}^h}  \frac{1}{\| \dispdiscr(t,\theta) ||^2_2} \left\| \dispdiscr(t,\theta) - \bar{\dispdiscr}(t,\theta) \right\|_2^2 \right) \, ,
\end{equation}
In the following, we will only work with the normalised functional, and therefore write $J$ for $J^{norm}$.

\subsection{Estimation of the predictive power of the empirical POD by re-sampling techniques}

We now need to evaluate the predictive power of the empirical POD model for an arbitrary random realisation $\theta \in \Theta$. This can be quantified by the expectation of the error of projection onto the empirical reduced space, which reads:
\begin{equation}
\label{eq:snapmetric2}
\tilde{J}(\basereduc^s) = E \left( \frac{1}{n_t} \,  \sum_{t \in \mathcal{T}^h}  \frac{1}{{\| \dispdiscr(t,\theta) ||^2_2}} \left\| \dispdiscr(t,\theta) - \basereduc^s \,  {\basereduc^s}^T \dispdiscr(t,\theta) \right\|_2^2  \right) \, .
\end{equation}
However, for the same reason invoked in the previous subsection, $\tilde{J}$ cannot be calculated, and needs to be estimated. The most straightforward estimator is $J^{s}$. However, $J^s$ is a strongly biased estimator of $\tilde{J}$ because $\basereduc^s$ was chosen as to minimise $J^s$. 
More precisely, the expectation of $J^{s}$ is lower than $\tilde{J}$, and it is therefore said to be an optimistic measure of the predictivity of the empirical POD. This can be explained qualitatively. The POD model is ``trained'' or ``fitted'' on the set of realisations $\Theta^s$. Therefore, the projection error corresponding to these particular realisations is lower than the projection error of a set of arbitrary realisations $\tilde{\Theta}^s \subset \Theta$. This type of behaviour is called over-fitting (or error of type III in statistics). One could then simply evaluate the statistical average of the error on ${\Theta}^s$ and report it as an unbiased estimate for $\tilde{J}$. This is called the holdout method, and is not usually favoured by statisticians as it requires to reserve realisations which cannot be used to increase the accuracy of the POD model.

A widely used technique to obtain an almost unbiased estimate of $\tilde{J}$ without the need of additional realisations is the cross-validation (see comprehensive reviews of re-sampling techniques in \cite{bradleygong1983}, and their application to principal component analysis in \cite{jackson1993,cangelosigoriely2007}). The idea is to simulate the prediction of errors on realisations that are independent of those used to fit the POD model (the so-called training set). To achieve this goal, we compute the error made successively on each of the realisations of the snapshot, while computing the POD with all the other realisations. We then sum up all the contributions to obtain the cross-validation estimate of the prediction error. For each of these contributions, an empirical reduced basis $\basereduc^{s-.}$ is computed without using the realisation whose contribution is currently being evaluated. This can be expressed mathematically as follows:
\begin{equation}
\label{eq:snapmetricCV}
\bar{J}^s(\basereduc^s) = \frac{1}{n_\theta} \, \frac{1}{n_t} \, \sum_{(t,\theta) \in \mathcal{P}^s}  \frac{1}{{\| \dispdiscr(t,\theta) ||^2_2}} \left\| \dispdiscr(t,\theta) - \basereduc^{s-\theta} \,  ({\basereduc^{s-\theta}})^T \dispdiscr(t,\theta) \right\|_2^2  \, .
\end{equation}
where $\basereduc^{s-\theta}$ is obtained for a particular realisation of the snapshot $\theta \in \Theta^s$ by minimisation of the subset POD functional
\begin{equation}
\label{eq:snapmetricCV2}
J^{s-\theta}_\theta(\basereduc^{s-\theta}) = \frac{1}{n_\theta-1} \, \frac{1}{n_t} \, \sum_{t \in \mathcal{T}^h, \, \tilde{\theta} \in \Theta^s \backslash \{\theta \} }  \frac{1}{{\| \dispdiscr(t, \tilde{\theta}) ||^2_2}} \left\| \dispdiscr(t,\tilde{\theta}) - \basereduc^{s-\theta} \,  ({\basereduc^{s-\theta}})^T \dispdiscr(t,\tilde{\theta}) \right\|_2^2  \, .
\end{equation}
$\bar{J}^s(\basereduc^s)$ is traditionally called PRESS (predicted residual sum of squares) in statistics. It can be shown to be slightly biased upward (pessimistic), as only $n_\theta-1$ of the elements of the snapshot are used to compute each of the individual projection errors. The version of PRESS given by equations (\ref{eq:snapmetricCV},\ref{eq:snapmetricCV2}) is called ``leave-one-out'' (LOOCV). Popular versions of the cross-validation, called n-folds, divide the realisation space $\Theta^s$ into $n$ subsets $\left( \Theta^s_n \right)_{n \in \llbracket 1 , \, ...  \, , n\rrbracket}$ of roughly identical cardinality. The cross-validation estimate $\bar{J}^s$ is then obtained as described previously, but removing a whole subset of elements of the snapshot to fit the successive empirical POD basis $\left( \basereduc^{s-}(\Theta^s_n) \right)_{n \in \llbracket 1 , \, ...  \, , n\rrbracket}$ used to calculate the elements of the sum over the realisations in \eqref{eq:snapmetricCV}. 

In our numerical experiments, We will use  the 10-fold CV when $n_\theta > 20$, and LOOCV otherwise. For a justification of this classical choice, the interested reader can refer to \cite{bradleygong1983}.

We define the error estimate provided by the cross-validation
\begin{equation}
  \bar{\nu}^s 
 = \sqrt{ \bar{J}(\basereduc^s) } .
\end{equation}



\subsection{Locally uncorrelated solutions in the case of stochastic fracture}
\label{sec:Locally_uncorrelate2}

\begin{figure}[htb]
       \centering
       \includegraphics[width=0.9 \linewidth]{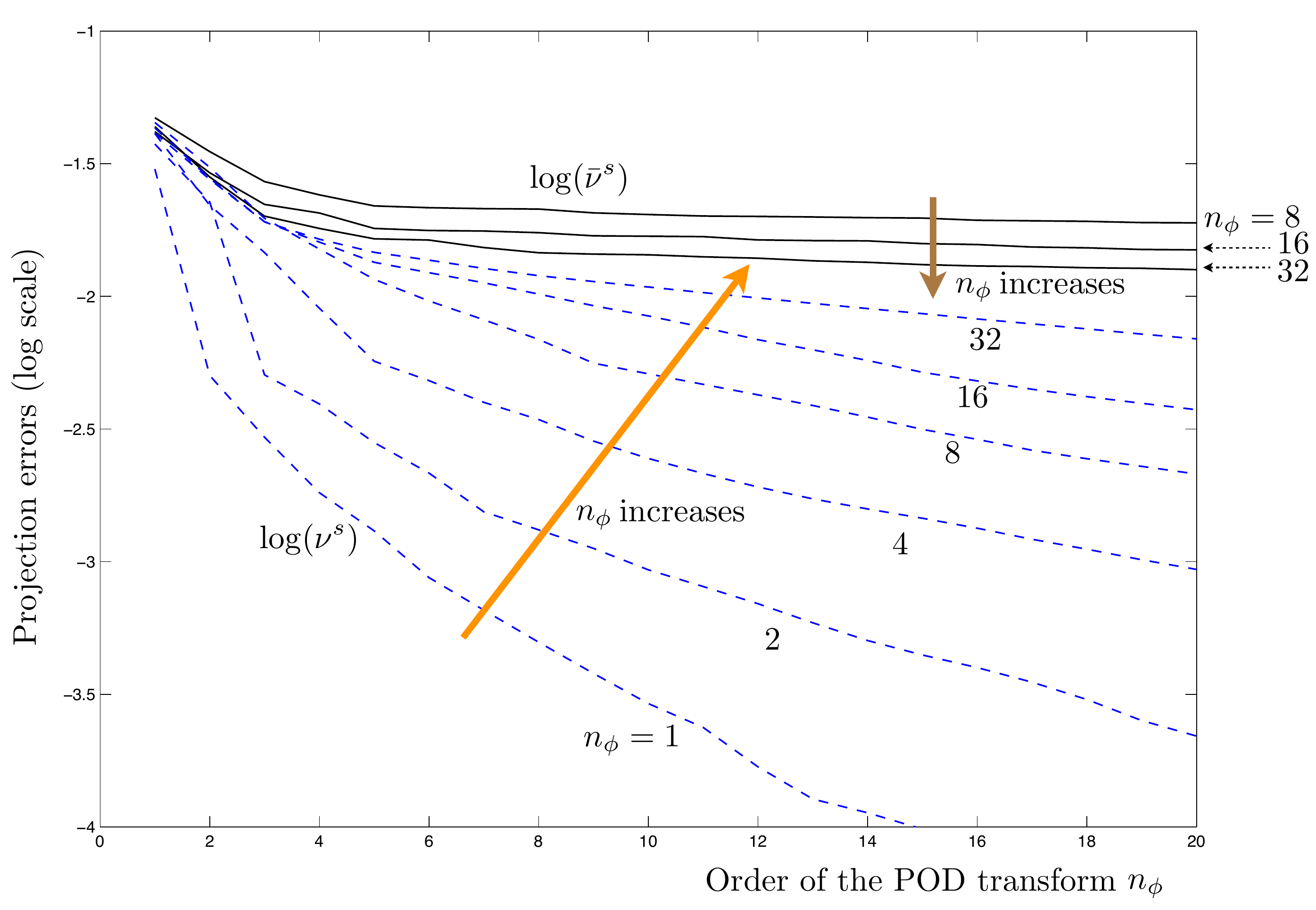}
       \caption{Statistical estimates of the error of projection of the empirical POD as a function of the order of the decomposition. The error curves in dashed line are evaluated directly on the samples used to obtain the POD model. The experiment is reproduced for different number of random realisations. The error curves in plain line are the convergence curve of the cross-validation estimate of the projection error as a function of the order of the decomposition, for snapshots of different cardinal.}
       \label{fig:ErrorUncorrelated}
\end{figure}



In the case of fracture mechanics, the predictive power of the empirical POD is too poor to be used in analysis or reduced order modelling. This is shown by the numerical results in figure \ref{fig:ErrorUncorrelated}. We show, in plain lines, the convergence of the cross-validated error estimate $\bar{\nu}^s$ as a function of the order of truncation of the POD projector, $n_\phi$. This is done for different sizes of sample space $n_\theta = 8$, $n_\theta = 16$ and $n_\theta = 32$. We can see that for a given order of the transform, the error estimate decreases slowly with the number of sampled realisations. More interestingly, each of the convergence curve decreases slowly for $n_\phi < 5$, and then flattens. The level of accuracy obtained is not satisfactory for the construction of a predictive model, where an accuracy of typically less than $10^{-3}$ is desired for each of the predicted realisations. Of course, $\bar{\nu}^s$ is a point estimate, and confidence intervals should be added to show the dependency of the results on the particular set of realisations used as snapshot. However, we only wish to show the overall behaviour of the POD transforms, and stating that similar trends have been observed with different sets of realisations of same cardinality is sufficient at this stage.

Further understanding can be provided by plotting the biased error estimates $\nu^s$ for different number of sampled realisations $n_\theta$. The corresponding convergence curves are plotted in dashed lines in figure\ref{fig:ErrorUncorrelated}. We can see that the behaviour of one particular realisation ($n_\theta = 1$) is captured with a high degree of accuracy $\nu^s \approx 10^{-3}$ with 5 or 6 spatial vectors. Therefore, the time response of the structure for a given distribution of the material properties is highly correlated, and can be approximated in a low-dimensional manifold of dimension 5. However, when performing the same experiment with two such realisations, the size of the reduced space required to approximate the two time evolutions is doubled. This behaviour still shows when further increasing the number of random samples, and leads to the flattening curves observed for the objective estimate $\bar{\nu}^s$. Notice that when $n_\theta$ tends to infinity, one should indeed see that $\nu^s$ tends to $\bar{\nu}^s$, providing that the bias of the cross-validation strategy is small enough. Physically speaking, the trends observed here suggest that each of the time evolutions of the random process requires its own reduced space, which is relatively independent on the space generated for the other ones. The stochastic fracture process is therefore uncorrelated. This is not surprising, looking at the various crack paths represented in figure \ref{fig:RandomCracks}. A subspace representing one of the time evolutions of the damage will not necessarily help predicting other crack paths.




\begin{figure}[htb]
       \centering
       \includegraphics[width=0.8 \linewidth]{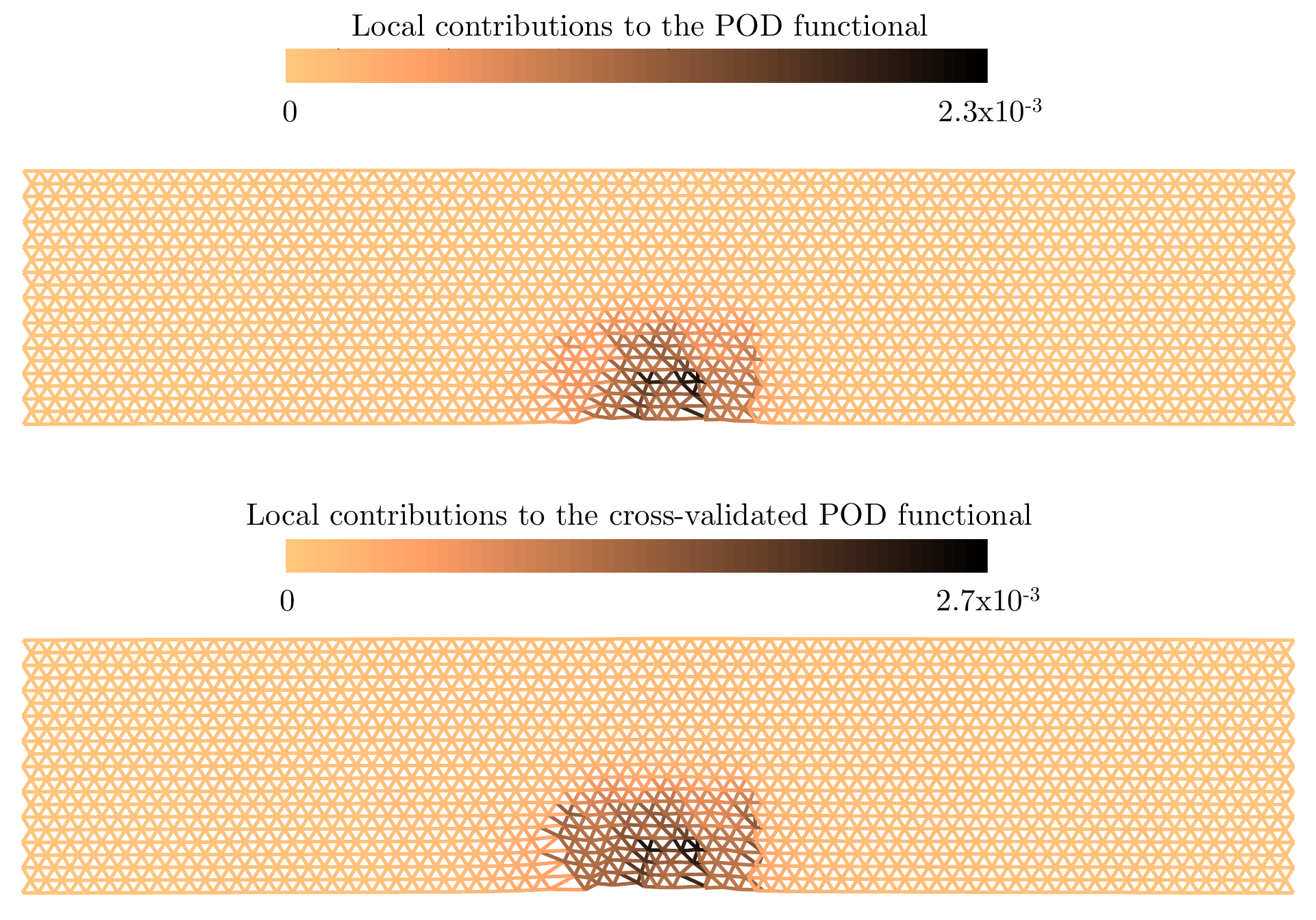}
        \caption{Local contributions to the error estimate of the empirical POD}
      \label{fig:LocalError}
\end{figure}


However, we can show that this issue comes from very localised uncorrelated data in space, in the region were the different cracks propagate. To show evidence of this fact, first notice that the functional that is minimised by the empirical POD transform \eqref{eq:snapmetricnorm} can be written as a sum of local contributions in space. Let us define the normalised difference between a time instance of an arbitrary realisation of the random process and its projection in the subspace generated by the empirical POD as:
\begin{equation}
\forall (t,\theta) \in \mathcal{P}, \quad \V{\boldsymbol{\epsilon}} (t,\theta) = \frac{1}{\| \dispdiscr(t,\theta) ||_2} \left( \dispdiscr(t,\theta) - \bar{\dispdiscr}^{s} (t,\theta) \right) \, .
\end{equation}
Then, equation \eqref{eq:snapmetricnorm} can be written as
\begin{equation}
\begin{array}{ll}
J^s(\basereduc^s) & \displaystyle  = \frac{1}{n_\theta} \, \frac{1}{n_t} \,  \sum_{(t,\theta) \in \mathcal{P}^s}  \V{\bm{\epsilon}}(t,\theta) ^T  \V{\bm{\epsilon}}(t,\theta) 
\\
& \displaystyle = \frac{1}{n_\theta} \, \frac{1}{n_t} \, \sum_{(t,\theta) \in \mathcal{P}^s} \sum_{i=1}^{n_u}  \bm{\epsilon}_i(t,\theta)^2
\\
& \displaystyle
= \sum_{i=1}^{n_u}  \left( \frac{1}{n_\theta} \, \frac{1}{n_t} \,  \sum_{(t,\theta) \in \mathcal{P}^s}   \bm{\epsilon}_i(t,\theta)^2 \right)
\end{array} \, .
\end{equation}
Defining vector the vector of local contributions to the empirical POD functional by $\V{\mathbf{Q}} \in \mathbb{R}^{n_u}$ by its components $\displaystyle \mathbf{Q}_i =  \left( \frac{1}{n_\theta} \, \frac{1}{n_t} \,  \sum_{(t,\theta) \in \mathcal{P}^s}   \bm{\epsilon}_i(t,\theta)^2 \right)^{1/2}$, where $i\in \llbracket 1, n_u\rrbracket$ we obtain
\begin{equation}
J^s(\basereduc^s) = \| \V{\mathbf{Q}} \|_2^2 \, .
\end{equation}
The same vector of local space contributions can be defined for the cross-validation estimate of the POD functional $\bar{J}^s$. Both vectors are represented in  figure \ref{fig:LocalError}. The deformation of the mesh  show the amplitude of the components of $\V{\mathbf{Q}}$ associated with each node of the lattice structure in the $\V{x}$ and $\V{y}$ directions, while the color scale represents the local contributions per node (square root of the sum of the square of the local contributions along $\V{x}$ and $\V{y}$). 16 random realisations have been used as a snapshot in the present case. The snapshot POD is truncated at order 5. It clearly appears that the major contributors to the empirical POD and empirical cross-validated POD functionals are localised in a narrow region: the process zone.



\subsection{Weighted POD to improve the convergence of POD transforms}

The weighted POD transforms (or more precisely POD transforms with a weighted inner product) are classically used to improve the convergence of the POD transform. A typical weighted version of the normalised POD defined previously minimises the following weighted distance:
\begin{equation}
 d^w_\mathcal{S}(\dispdiscr,\bar{\dispdiscr}) = \sum_{\theta \in \Theta} \frac{P(\theta)}{n_t} \sum_{t \in \mathcal{T}^h}  \frac{\omega(t,\theta)}{\| \dispdiscr(t,\theta) ||^2_2} \| \dispdiscr(t,\theta) - \bar{\dispdiscr}(t,\theta) \|^2_{\M{\mathbf{L}}} \, ,
 \end{equation}
where $\forall (t,\theta) \in \mathcal{P}, \,  \bar{\dispdiscr} = \basereduc \, \coeffreduc(t,\theta)$, and the weighted space inner product is defined by $\| \X \|_{\M{\mathbf{L}}} = \sqrt{\X^T \M{\mathbf{L}} \, \X} $, with $\M{\mathbf{L}}$ a diagonal, definite positive matrix. Adjusting space weights $\M{\mathbf{L}}$ may permit to retrieve a fast convergence in space when locally uncorrelated data prevents it, which is of particular interest to us. Similarly, the scalar weight function $\omega$ defined over $\mathcal{P}$ might allow to enhance the convergence properties of the POD when locally uncorrelated time instances or random realisations prevents the construction of a low-dimensional attractive subspace.

A weighted version of the empirical POD can also be defined by minimising functional
\begin{equation}
J^{s,w}(\basereduc^s) = \frac{1}{n_\theta} \, \frac{1}{n_t} \sum_{(t,\theta) \in \mathcal{P}^s} \frac{\omega(t,\theta)}{\| \dispdiscr(t,\theta) ||^2_2} \| \dispdiscr(t,\theta) - \basereduc^s {\basereduc^s}^T {\M{\mathbf{L}}} \, \dispdiscr(t,\theta) \|^2_{\M{\mathbf{L}}} 
 \end{equation}
In addition to the previous remarks on the potential usage of these weights, $\omega$ may allow to reduce the effect of outliers (extreme values of the random process) from the analysis, by an approximation of their respective probability, therefore allowing the empirical POD to converge faster towards the POD in terms of size of sample space. Although this particular feature is also of interest to us, we focus in this work on space weights ${\M{\mathbf{L}}} $.

 Intuitively, we would like to set the entries of $\M{\mathbf{L}}$ that correspond to the process zone to very small values. How small these weights should be is obviously a difficult question to address. Too small weights would lead to numerical ill-posedness of the POD minimisation problem, while too large ones would pollute the results in the zones where the random process is correlated. We propose to extend the idea of weighted POD transforms by performing a POD decomposition of the data corresponding to only a subset of the spatial degrees of freedom. This set of degrees of freedom, called spatial domain of validity of the POD transform, are the degrees of freedom for which the local contributions to the POD metric $\left( \mathbf{Q}_i \right)_{i \in \llbracket 1, n_u \rrbracket}$ is small enough (broadly, the space components for which the POD ``works''). In turn, obtaining these local contributions requires the knowledge of an initial POD transform. Hence, we arrive at the idea of an iterative process to find an optimal POD-type decomposition associated with a domain of validity, which is in substance what is proposed in the following section.



\section{Progressive Restricted POD}


\subsection{Limit case of Weighted POD: the Restricted POD}
\label{sec:RestrictedPOD}

The proposed restricted POD performs a classical POD of the spatial components of the snapshot samples corresponding to a high level of correlation (i.e. away from the process zone). These components will be denoted by superscript $^{(r)}$. This transform being computed, a POD-type decomposition of the remaining components (corresponding to the damage zone), denoted by $^{(f)}$ can obtained by solving a complementary minimisation problem (i.e.: a regularisation of the weighted POD problem).

Formally, let us decompose the euclidean space $\mathbb{R}^{n_u}$ as a product of two spaces: $ \mathbb{R}^{n_u} = \mathbb{R}^{n_r} \times \mathbb{R}^{n_f}$, with $n_u=n_r+n_f$. Any vector $\X \in \mathbb{R}^{n_u}$ will be decomposed as follows
\begin{equation}
\X = \left( \X^{(r)}, \X^{(f)} \right) \in \mathbb{R}^{n_r} \times \mathbb{R}^{n_f} \, .
\end{equation}
We introduce the boolean restriction operators $\M{\mathbf{E}}^{r} \in  \{ 0 , 1 \}^{n_r \times n_u}$ and $\M{\mathbf{E}}^{f} \in \{ 0 , 1 \}^{n_f \times n_u}$ as
\begin{equation}
\begin{array}{c}
\displaystyle \X^{(r)} = \M{\mathbf{E}}^{r} \X \\
\displaystyle \X^{(f)} = \M{\mathbf{E}}^{f} \X
\end{array} \, ,
\end{equation}
Only one entry per line of rectangular operators $\M{\mathbf{E}}^{r}$ and $\M{\mathbf{E}}^{f}$ is equal to one; the remaining entries are null. Hence, $\X \in \mathbb{R}^{n_u}$ can be written as a sum of two contributions
\begin{equation}
\X  = {\M{\mathbf{E}}^{r}}^T \X^{(r)} + {\M{\mathbf{E}}^{f}}^T \X^{(f)} \, .
\end{equation}
Hence, ${\M{\mathbf{P}}^{r}} = {\M{\mathbf{E}}^{r}}^T {\M{\mathbf{E}}^{r}}$ is a diagonal orthogonal projector which, when applied to arbitrary vector $\X$, sets the entries of $\X$ corresponding to the process zone to $0$, while the entries of $\X$ that correspond to the domain of validity of the POD are unchanged. A similar remark leads to the definition of projector ${\M{\mathbf{P}}^{f}} = {\M{\mathbf{E}}^{f}}^T {\M{\mathbf{E}}^{f}}$ and we have ${\M{\mathbf{P}}^{f}} + {\M{\mathbf{P}}^{r}} = \M{\mathbf{I}}_d$.

We directly describe the restricted POD in its empirical version, the deterministic counterpart being obtained by simply replacing the sum over the samples by an expectation in the distance expression given below. We therefore look for a decomposition of the sample realisation of the random process in the form
\begin{equation}
\label{eq:SepVariablesDiscrTimeRestr}
\begin{array}{crcl}
\displaystyle  \bar{\V{\mathbf{\mathbf{U}}}}^s  : & \  \mathcal{P}^s & \displaystyle \rightarrow & \displaystyle \mathbb{R}^{n_u} 
\\ \displaystyle
& \displaystyle (t,\theta) &  \mapsto & \displaystyle \left(  {\M{\mathbf{E}}^{r}}^T \basereduc^{s,(r)} + {\M{\mathbf{E}}^{f}}^T \basereduc^{s,(f)} \right) \coeffreduc^s(t,\theta) \, ,
\end{array}
\end{equation}
where $\coeffreduc^s(t,\theta)$ is a vector of weight functions depending on time and realisation index $\theta$ of the random process. Our goal is to find a basis $\basereduc^s$, and a restriction operator ${\M{\mathbf{E}}^{r}}$, which minimise a certain cost functional. To arrive at the formal expression of such a problem, let us first assume that an optimal splitting of $\mathbb{R}^{n_u}$ is known. The restriction of the transform to the $^{(r)}$ degrees of freedom $\bar{\V{\mathbf{\mathbf{U}}}}^{s,(r)}$ is then required to minimise a distance with respect to $\V{\mathbf{\mathbf{U}}}^{s,(r)}$, given by 
\begin{equation}
\label{eq:restricted_transform}
d^{s,r}(\dispdiscr^{(r)},\bar{\dispdiscr}^{s,(r)}) =   \frac{1}{n_\theta} \,    \frac{1}{n_t} \sum_{(t,\theta) \in \mathcal{P}^s} \frac{1}{\| \dispdiscr^{(r)}(t,\theta) \|_2^2} \| \dispdiscr^{(r)}(t,\theta) - \bar{\dispdiscr}^{s,(r)}(t,\theta)\|_2^2 \, .
\end{equation}
The optimisation is constrained by the orthonormality condition
\begin{equation}
{\basereduc^{s,(r)}}^T {\basereduc^{s,(r)}} = \M{\mathbf{I}}_d
\end{equation}
This transform is  a classical empirical (normalised) POD in $\mathbb{R}^{n_r}$. Its solution is as follows:
\begin{itemize}
\item  $( \V{\boldsymbol{\phi}}_i^{s,(r)} )_{i \in \llbracket 1, n_\phi \rrbracket}$ are the eigenvectors of $\displaystyle \M{\mathbf{H}}^r= \frac{1}{n_\theta} \,    \frac{1}{n_t} \sum_{(t,\theta) \in \mathcal{P}^s} \frac{1}{\| \dispdiscr^{(r)}(t,\theta) \|_2^2} {\M{\mathbf{E}}^{r}} \dispdiscr(t,\theta) \, \dispdiscr(t,\theta)^T {\M{\mathbf{E}}^{r}}^{T} $ associated to its largest $n_\phi$ eigenvalues $(\lambda_i)_{i \in \llbracket 1, n_\phi \rrbracket}$.
\item $\forall (t,\theta) \in \mathcal{P}^s$,  $\coeffreduc^s(t,\theta) =  {\basereduc^s}^T   {\M{\mathbf{E}}^{r}}^{T}  {\M{\mathbf{E}}^{r}} \, \V{\mathbf{\mathbf{U}}}(t,\theta) $ (the weights are not defined by the empirical transform for $\theta \notin \Theta^s$). $\coeffreduc^s$ is completely determined, for any $\theta \in \Theta^s$, by the minimisation problem on the $^{(r)}$ degrees of freedom.  Hence we will use hereafter the notation $\coeffreduc^{s,r} = \coeffreduc^s$.
\end{itemize}

We will from now on use the notation $\nu^{s,r} = \sqrt{d^{s,r}(\dispdiscr^{(r)},\bar{\dispdiscr}^{s,(r)})}$ for the error estimate corresponding to this transform. As for usual POD transforms, this estimate is the square root of the sum of the ordered eigenvalues of $\M{\mathbf{H}}^r$ with indices larger than the order of truncation $n_\phi$. 

We now discuss how to obtain the complementary part of the decomposition $\bar{\dispdiscr}^{s,(f)}$. This is not a fundamental ingredient of the statistical extraction, as the previous decomposition associated to a given partitioning of the degrees of freedom can be performed in a ``stand-alone'' manner. However, a prolongation of the reduced basis to the process zone can be useful when the aim is to define a global preconditioner, as in our previous investigations on the subject for instance \cite{kerfridengosselet2010,kerfridenpassieux2011}. If necessary, the complementary part of the decomposition may be obtained by minimisation of the following functional:
\begin{equation}
J^{s,f} ( \basereduc^{s,(f)} ) = \frac{1}{n_\theta} \,    \frac{1}{n_t} \sum_{(t,\theta) \in \mathcal{P}^s} \frac{1}{\| \dispdiscr^{(f)}(t,\theta) \|_2^2} \| \dispdiscr^{(f)}(t,\theta) - {\basereduc^{s,(f)}} \, \coeffreduc^{s,r}(t,\theta) \|_2^2 \, .
\end{equation}
The previous problem can be interpreted as follows: we look for a prolongation $\dispdiscr^{(f)}$ of the basis $\dispdiscr^{(r)}$ such that if the projector is fitted on the reduced degrees of freedom, the global projection error (comprising the process zone) is minimal. The solution to this problem is given by:
\begin{equation}
\basereduc^{s,(f)}  =  \M{\mathbf{G}} \, \M {\bm \gamma}^{-1} \, ,
\end{equation}
where $\displaystyle  \M{\mathbf{G}} = \left( {\M{\mathbf{E}}^{f}} \sum_{(t,\theta) \in \mathcal{P}^s}  \dispdiscr(t,\theta) \,  \coeffreduc^{s,r}(t,\theta)^{T} \right)$ 
and $\displaystyle \M {\bm \gamma} = \sum_{(t,\theta) \in \mathcal{P}^s} \coeffreduc^{s,r}(t,\theta) \, \coeffreduc^{s,r}(t,\theta)^{T} $

Solving the complementary minimisation problem can be seen as a regularisation of a classical weighted POD, when the space weights $\M{\mathbf{L}}$ are $1$ for the $^{(r)}$ degrees of freedom, and $0$ for the $^{(f)}$ remaining ones. An error estimate can be defined for this part of the solution, but is not used in the following, as by construction, the $^{(f)}$ degrees of freedom carry most of the error, and the corresponding reduced basis cannot be used solely as a predictor.





We are now able to express the problem of the restricted POD in full. We look for a transform of type \eqref{eq:restricted_transform} which minimises the associated error estimate $\nu^{s,r}$, under the constraint that the extractor $\M{\mathbf{E}}^{r}$ is of given rank $n_r$ (in other words, the size of the process zone, characterised by $n_f=n_u-n_r$, is fixed in advance). In the following section, we propose a greedy algorithm, which provides a suboptimal solution to this problem. The choice of the best $n_r$ for our practical application is then obtained by a post-treatment of a range of solutions of the restricted POD corresponding to various $n_r$.

\subsection{Computation of Restricted POD and associated validity domain with a greedy algorithm}
\label{sec:ProgressivePOD}

\subsubsection{General algorithm}



The Greedy algorithm proposed in this section can be viewed as an element of a general class of iterative algorithms in two steps. We first solve the POD problem, given the functional to minimise (i.e.: the weights the restriction in our case). We then update the parameters of the functional such that the decrease in the error in the POD model is as large as possible. This second step is performed using the projector given by the first step, and under a set of constraints defining how these parameters can be updated at each iteration. The iterative algorithm is stopped when a certain target, or a certain minimum, is achieved.

Let us now apply this general concept to the restricted POD described in the previous section. We introduce the functional $\hat{J}^{s,r}( \basereduc^{s}, \, \M{\mathbf{E}}^{r}) = d^{s,r}(\dispdiscr^{(r)},\bar{\dispdiscr}^{s,(r)})$, where $\bar{\dispdiscr}^{s,(r)} = {\basereduc^{s,r}}^T \, \basereduc^{s,r} \, \dispdiscr^{(r)} $, and for a given order $n_\phi$ of the decomposition. The greedy optimum $\left( \basereduc^{s}, \M{\mathbf{E}}^{r} \right)$ is obtained in the following way:
\begin{enumerate}
\setcounter{enumi}{-1}
\item \textit{Initialisation}. Initialise the algorithm with $\M{\mathbf{E}}^{r}_{ \, 0} = \M{\mathbf{I}}_d$, where $\M{\mathbf{I}}_d$ is the identity operator in $\mathbb{R}^{n_u \times n_u}$, and set $i=1$
\item \textit{POD}. Find a reduced basis $\basereduc^{s,r}_{\, i}$ that minimises $\hat{J}^{s,r}( \basereduc^{s,r}_{\, i} , \, \M{\mathbf{E}}^{r}_{ \, i-1} )$, and compute the prolongation $\basereduc^{s,f}_{\, i}$ by minimising the complementary cost function $\hat{J}^{s,f}( \basereduc^{s,f}_{\, i} , \, \M{\mathbf{E}}^{f}_{ \, i-1} ) = J^{s,f}( \basereduc^{s,f}_{\, i} )$, where $J^{s,f}( \basereduc^{s,f}_{\, i} )$ is evaluated for fixed extractor $\M{\mathbf{E}}^{f}_{ \, i-1}  $.
\item \textit{Stopping criterion}. If the stopping criterion is reached stop the greedy algorithm and return suboptimal solution $\left( \basereduc_{\, i}^{s}, \M{\mathbf{E}}_{\, i-1}^{r} \right)$. 
\item \textit{Update of the POD functional}. Compute an update  $\M{\Delta \mathbf{P}}^{r}_{ \, i} =\M{ \mathbf{P}}^{r}_{ \, i} - \M{ \mathbf{P}}^{r}_{ \, i-1}$ of projector $\mathbf{P}^{r}$, such that 
the rank of $\M{\mathbf{E}}^{r}$ is reduced and 
$\hat{J}^{s,r}( \basereduc^{s}_{\, i} , \, \M{\mathbf{E}}^{r}_{ \, i} )$ is minimum under some constraints on the admissibility of the parameter update.
\item $i \leftarrow i+1$.
\end{enumerate}

Step 2 has been defined in the previous subsection. Our choice for the initialisation of the algorithm is to perform a classical POD on the whole spatial domain. We now need to clarify step 3 and 4 of the algorithm.

\subsubsection{Update of the extractors (Step 4 of the Greedy algorithm)}

At iteration $i$ of the greedy algorithm, we first look here for an optimal update $\M{\Delta \mathbf{P}}^{r}_{ \, i}$ of rank $1$ of the POD functional. In other words, we set one non-zero entry of the $(i-1)^{\textrm{th}}$ iterate of projector ${\mathbf{P}^{r}}$ to 0. This entry, indexed $j$, corresponds to the maximum spatial contribution of the POD error, which formally reads
\begin{equation}
\displaystyle j = \underset{j \in \llbracket 1, n_u \rrbracket}{\argmax} \, \left( \sum_{(t,\theta) \in \mathcal{P}^s} \bm{\epsilon}^2_j (t,\theta) \right) \ ,
\end{equation}
where $\forall(t,\theta)\in\mathcal{P}, \ \bm{\epsilon}_j(t,\theta)$ is the $j^{\textrm{th}}$ component of residual vector:
\begin{equation}
\V{\bm{\epsilon}}(t,\theta) = 
\frac{1}{\| \M{\mathbf{E}}^{r}_{ \, i-1} \dispdiscr(t,\theta) \|_2} \, 
\left({\M{\mathbf{E}}^{r}_{ \, i-1}} \right)^T \M{\mathbf{E}}^{r}_{ \, i-1} \left( \M{\mathbf{I}}_d - \basereduc^s_{ \, i} \, \left( \basereduc^s_{ \, i} \right)^T  \, \left( {\M{\mathbf{E}}^{r}_{ \, i-1}}  \right)^T \, \M{\mathbf{E}}^{r}_{ \, i-1} \right)  \dispdiscr(t,\theta) 
\end{equation}
We have here made use of the work explained previously, which allowed us to rewrite the POD error as a sum of local contributions over space. 

Notice that in order to find index $j$ of the space degree of freedom to be added to the process zone, the order of the POD (the rank of $\basereduc$), needs to be defined. In our implementation of the strategy, we perform separate greedy computations for $n_\phi$ ranging from $1$ to a value specified by the user, and try to extract a good (or optimal in some sense) value of $n_\phi$ in a post-treatment phase.

Looking for a rank-one update of the projector $\M{\mathbf{P}}^{r}$ can be computationally expensive if the process zone contains a large number of degrees of freedom. Therefore, in a second stage of the update, we set to zero all the entries of $\M{\mathbf{P}}^{r}$ that correspond to nodes of the lattice structure which are located in a sphere on radius $\rho$ centred on the node carrying entry $j$. This particular feature also permit to obtain a compact modification of the process zone at each step of the greedy algorithm.

\subsubsection{Stopping criterion}

We simply stop the algorithm when the rank of the $i^\textrm{th}$ iterate of projector $\M{\mathbf{P}}^{r}$ is smaller than a value specified by the user. As for the choice of the best truncation order of the POD, the extraction of an``optimal'' rank of the projector is done in a post-treatment phase which shall be described later on.

\subsubsection{Illustration}

\begin{figure}[htb]
       \centering
       \includegraphics[width=0.8 \linewidth]{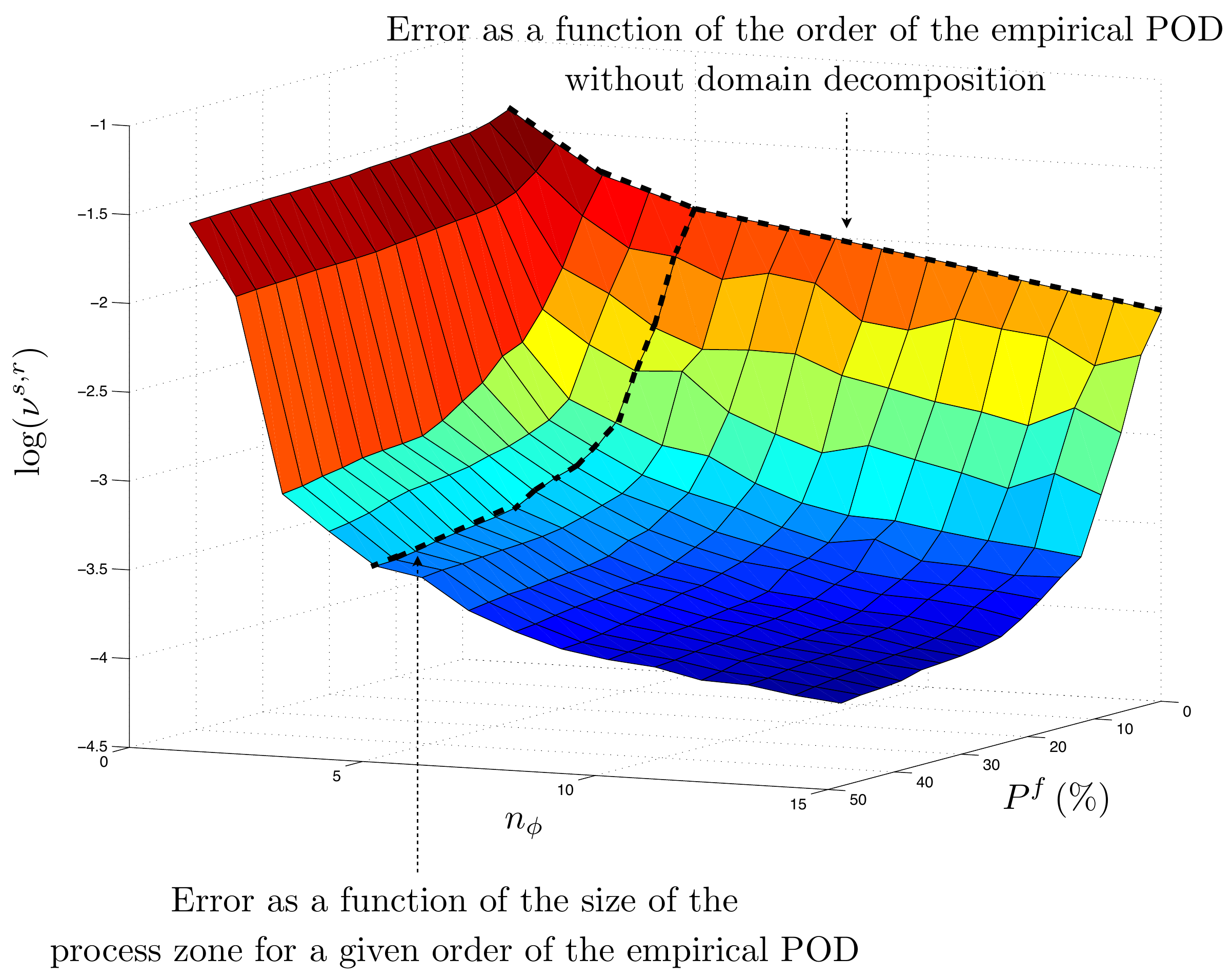}
       \caption{Estimate of the prediction error of the restricted POD model as a function of the number of modes the number of spatial degrees of freedom over which the spectral decomposition is not performed (i.e. the ''size'' of the process zone)}
       \label{fig:ErrorPOD16}
\end{figure}

Let us illustrate the ideas presented  in this section by analysing the results given in figure \ref{fig:ErrorPOD16}. This figure shows the value of the empirical error estimate $\nu^{s,r}$ as a function of both the order $n_\phi$ of the decomposition and the rank of $\M{\mathbf{P}}^{f}$. For each grid point of the surface represented in figure \ref{fig:ErrorPOD16}, a greedy computation as been performed to obtain both the POD projection and the associated domain of validity. We used a snapshot containing $n_\theta=16$ realisations of the random process of fracture. The rank of $\M{\mathbf{P}}^{f}$ is described in a normalised form, 
\begin{equation}
P^f = \frac{\textrm{rank} \left( \M{\mathbf{P}}^{f} \right)}{n_u} \, .
\end{equation}
As the mesh of the lattice structure is regular, $P^f$ approximates the ratio between the surface area of the process zone and the total surface of the structure. The curve corresponding to $P^f=0$ is the one displayed in figure \ref{fig:ErrorUncorrelated} for $n_\theta=16$ (empirical POD over all the unknowns).

Two general tendencies appear in this graph. The error estimate $\nu^{s,r}$ decreases with increasing relative size of the process zones $P^f$, and with the order of the POD $n_\phi$. Notice that for a fixed size of the process zone, the error does not necessarily decrease monotonically. Indeed, the domains of validity of the associated solutions, although of fixed size, might differ. Therefore, one cannot apply the classical results on POD transforms, as two successive solutions do not necessarily work on the same data. However, due to the definition of the greedy algorithm given previously, the error does decrease monotonically  with an increasing size of the process zone, for a fixed order of  the POD approximation.

In more details, one can see that providing that the order of the decomposition is sufficiently large to allow for the extraction of a good spatial basis for the random process, the error first decreases quickly when increasing $P^f$ (between around 0 and 10 \% in this example), and then continues to decrease but at a lower rate. Conversely, providing that we remove a sufficiently large part of the uncorrelated spatial degrees of freedom to perform the POD, the error decreases quickly with the order of truncation, and then tends to stagnate (around $n_\phi=5$ in this example). This indicates that:
\begin{itemize}
\item providing that the process zone is excluded from the correlation analysis, the spatial dimensionality of the attractive subspace of the problem of stochastic fracture is low and relatively well-defined. It also leads to an average prediction error which is acceptable for an engineering application (around $10^{-3}$).
\item The effect of the random crack paths is relatively localised. A small number of POD modes (smaller than the number of realisations of the process) permit to represent the random process with a high level of fidelity away from a relatively narrow region.
\item The definition of the process zone is not obvious. The error decreases continuously with $P^f$.
\end{itemize}


We took some precautions when stating the previous conclusions, as the analysis is in fact restricted to the particular sample which has been used to construct the reduced spaces and associated domain of validity. In fact, $\nu^{s,r}$ is an optimistic measure of the predictive behaviour of a model constructed by the empirical restricted POD (it is biased downwards). Therefore, a re-sampling strategy needs to be defined in order to produce an objective confirmation of these observations.



\subsection{Cross-validation of the restricted POD}

\begin{figure}[htb]
       \centering
       \includegraphics[width=0.8 \linewidth]{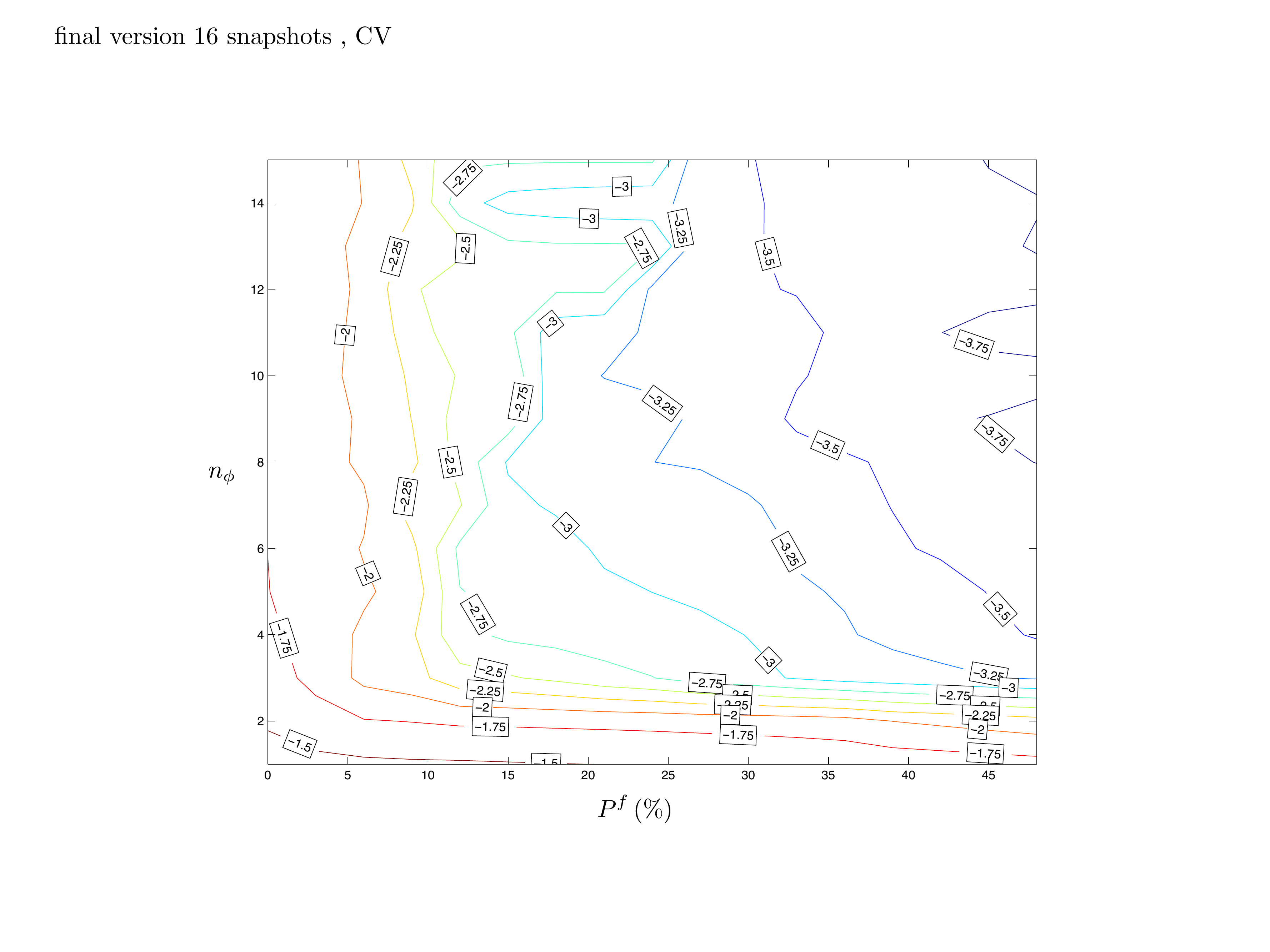}
       \caption{Cross-validation estimate of the projection error of the restricted POD}
       \label{fig:ErrorPOD16CV}
\end{figure}

The cross-validation estimate $\bar{\nu}^{s,r}$ of the restricted POD is an almost unbiased estimate for the expectation of the prediction error made when using the empirical restricted POD model
\begin{equation}
\left( \tilde{\nu}^{s,r} \right)^2 = E \left( \,    \frac{1}{n_t} \sum_{t \in \mathcal{T}^h} \frac{1}{\| \dispdiscr^{(r)}(t,\theta) \|_2^2} 
\left\| \dispdiscr^{(r)}(t,\theta) -  {\basereduc^{s,(r)}} \, {\basereduc^{s,(r)}}^T \dispdiscr^{(r)}(t,\theta) \right\|_2^2  \right)  \, .
\end{equation}
An important feature needs to be highlighted here. Let us recall that the cross validation simulates the prediction of independent realisations by the empirical model. In order to achieve this, we ignore, in turn, each of the observations of the snapshot from the set used to fit the POD model. We fit the model on the amputated snapshot, and measure the error made when predicting the realisation which has been ignored. All the individual errors are then summed up to obtain the cross-validation estimate. In the case of the restricted POD, building the model means extracting an attractive spatial manifold of specified dimension, and an associated domain of validity. Therefore, in order to obtain an objective measure of the predictive ability of the empirical restricted model, we need take into account this remark in both the POD step and the update step of the greedy algorithm. More precisely,
\begin{equation}
\begin{array}{rl}
\displaystyle \left( \bar{\nu}^{s,r} \right)^2 = & \displaystyle \frac{1}{n_\theta} \,    \frac{1}{n_t} \sum_{(t,\theta) \in \mathcal{P}^s} 
\left( \phantom{\frac{a}{b}} \right.
\frac{1}{\|  \M{\mathbf{E}}^{r-\theta}_{ \, i-1}  \, \dispdiscr(t,\theta) \|_2^2}  \, 
\\
& \displaystyle \times \left\| 
\M{\mathbf{E}}^{r-\theta}_{ \, i-1} \,  \dispdiscr(t,\theta) -   {\basereduc_{ \, i}^{s-\theta,(r)}} \, \left( {\basereduc_{ \, i}^{s-\theta,(r)}} \right)^T \M{\mathbf{E}}^{r-\theta}_{ \, i-1} \, \dispdiscr(t,\theta) 
\right\|_2^2  \left. \phantom{\frac{a}{b}} \right)  \, .
\end{array}
\end{equation}
where ${\basereduc_{ \, i}^{s-\theta,(r)}} = \M{\mathbf{E}}^{r-\theta}_{ \, i-1} \, \basereduc_{ \, i}^{s-\theta}$ minimises
\begin{equation}
\begin{array}{rl}
\displaystyle J^{s-\theta,r}_\theta ({\basereduc_{ \, i}^{s-\theta,(r)}} ) = & \displaystyle \frac{1}{n_\theta-1} \,    \frac{1}{n_t} \sum_{t \in \mathcal{T}^h, \, \tilde{\theta} \in \Theta^s \backslash \theta} 
\left( \phantom{\frac{a}{b}} \right.
\frac{1}{\|  \M{\mathbf{E}}^{r-\theta}_{ \, i-1} \,   \dispdiscr(t,\tilde{\theta}) \|_2^2} \, 
\\ & \displaystyle \times  \left\|
\M{\mathbf{E}}^{r-\theta}_{ \, i-1} \, \dispdiscr(t,\tilde{\theta}) -   {\basereduc_{ \, i}^{s-\theta,(r)}} \, \left( {\basereduc_{ \, i}^{s-\theta,(r)}} \right)^T \M{\mathbf{E}}^{r-\theta}_{ \, i-1} \, \dispdiscr(t,\tilde{\theta}) 
\right\|^2 \left. \phantom{\frac{a}{b}} \right)
\end{array}
\end{equation}
Throughout the greedy process, the extractor $\M{\mathbf{E}}^{r-\theta}$ is updated without using the realisation of the random process corresponding to random distribution $\theta$ of the material properties. At iteration $i$, $\M{\mathbf{E}}_{\, i}^{r-\theta}$ is therefore updated from its previous iterate $\M{\mathbf{E}}_{\, i-1}^{r-\theta}$ by deleting line j of $\M{\mathbf{E}}_{\, i-1}^{r-\theta}$ (and the lines corresponding to nodes located in a sphere centred on the node carrying space degree of freedom $j$), where 
\begin{equation}
j = \underset{j \in \llbracket 1, n_u \rrbracket}{\textrm{argmax}} \left( \sum_{t \in \mathcal{T}^h, \, \tilde{\theta} \in \Theta^s \backslash \{ \theta \} } \left( {\bm{\epsilon}_{j}^{-\theta}} (t,\tilde{\theta}) \right)^2 \right)
\end{equation}
where $\bm{\epsilon}_{j}^{-\theta}$ is the $j^{\textrm{th}}$ component of residual vector:
\begin{equation}
\begin{array}{rl}
\displaystyle \V{\bm{\epsilon}}^{-\theta}(t,\tilde{\theta}) = & \displaystyle
\frac{1}{\| \M{\mathbf{E}}^{r-\theta}_{ \, i-1} \, \dispdiscr(t,\tilde{\theta}) \|_2} \, 
 {\M{\mathbf{E}}^{r-\theta}_{ \, i-1} }^T \M{\mathbf{E}}^{r-\theta}_{ \, i-1}
 \\ & \displaystyle \times 
 \left( \M{\mathbf{I}}_d - \basereduc^{s-\theta}_{ \, i}  \, \left( \basereduc^{s-\theta}_{ \, i} \right)^T  \, \left( {\M{\mathbf{E}}^{r-\theta}_{ \, i-1}} \right)^T \, \M{\mathbf{E}}^{r-\theta}_{ \, i-1} \right)  \dispdiscr(t,\tilde{\theta}) 
 \end{array}
\end{equation}
The cross-validation estimate $\tilde{\nu}^{s,f}$ is relatively tedious to implement, and the technical aspects will not be detailed any further. Just notice that its computation requires, by induction, to perform  the greedy algorithm  separately for each of the realisations that are successively removed from the training set of samples, and only then sum the contributions of the successively ignored data to the error estimate.


In figure \ref{fig:ErrorPOD16CV}, we plot the cross-validated estimate $\bar{\nu}^{r}$ obtained when using the cross-validated greedy algorithm to construct a restricted POD model. Notice first that the trends described for the evolution of $\nu^{s,r}$ as a function of $P^r$ and $n_\phi$ are still valid for $\bar{\nu}^{r}$. However, the flattening effect as the order of the decomposition increases is more acute, as the cross-validated estimate does not overfit the samples when $n_\phi$ becomes large. More importantly, when the size of the process zone is relatively small, and the order of the decomposition gets large, the cross-validation estimate of the projection error increases. This effect is well-known to statisticians. The cross-validation error estimate is indeed an (almost) unbiased estimate for $\bar{\nu}^{s,r}$. However, the process zone and complementary reduced space are both identified on the same samples. These two quantities tends to be arranged, through the minimisation process, in a way that minimises the projection error over these particular samples. Hence, an other kind of over-fitting behaviour is obtained, which usually results in a low convergence of the solution with the number of realisations.

\subsection{Double Cross-validation of the Restricted POD}

To avoid the inconvenience mentioned in the previous subsection, the usual remedy is to perform a double cross-validation \cite{wold1978}. The double cross-validation estimate proposes to couple the cross-validation and simple holdout methods. The sample space $\Theta^s$ is split into two complementary groups $\Theta_\phi^s$ and $\Theta_E^s$ (of roughly equal cardinality in our implementation). In the first step of the cross-validation greedy algorithm, the projector is identified using only the realisations corresponding to sample subspace $\Theta_\phi^s$. Similarly, in the step of update of the process zone, the projection error used to find a maximum local contribution is only computed using the realisations corresponding to random distributions in $\Theta_E^s$.


\begin{figure}[htb]
       \centering
       \includegraphics[width=0.8 \linewidth]{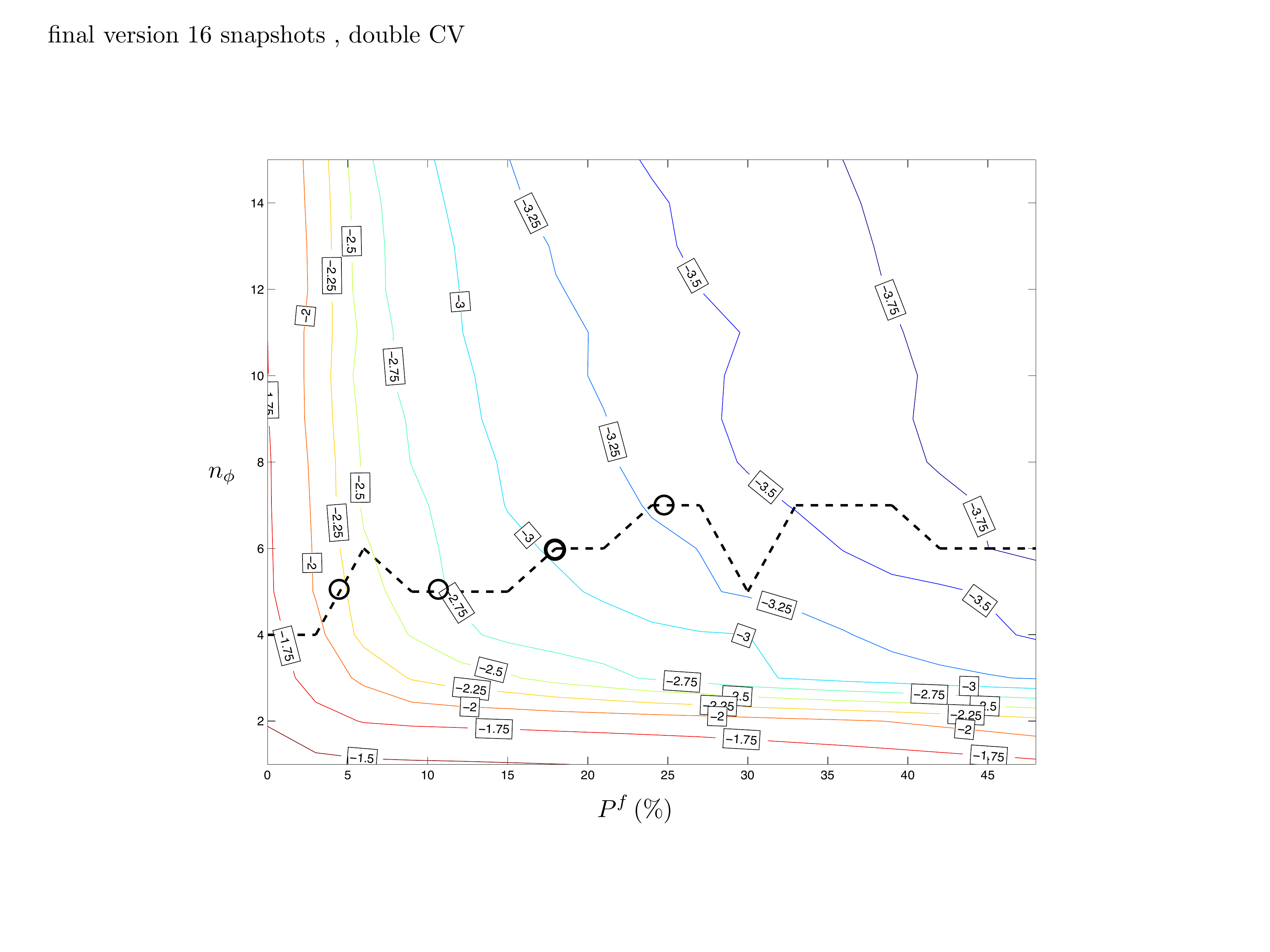}
       \caption{Double cross-validation estimate of the projection error of the restricted POD. The dashed line is the number of modes that are selected for different size of the process zone. The process zones obtained at the circled points are the one represented in figure \ref{fig:ProcessZone}.}
       \label{fig:ErrorPOD16doubleCV}
\end{figure}

The result of this procedure is given in figure \ref{fig:ErrorPOD16doubleCV}. The snapshot comprising $n_\theta=16$ time realisations of the random process has been randomly split into two groups $\Theta_\phi^s$ and $\Theta_E^s$, each composed of 8 distinct elements. One can see that for small sizes of the process zone, the level of error achieved with the double cross-validation technique is much lower than with the ``single'' cross-validation.

We now are in possession of a very useful tool for the analysis of random process with locally uncorrelated data, given a set of representative solution samples. These results need now be post-treated to select an appropriate extractor of the space degrees of freedom corresponding to correlated data, together with a projector of rank $n_{\phi}$ onto a representative reduced space. Further analysis would be required to assess whether a sufficiently large number of samples have been pre-computed in order to: 
\begin{itemize}
\item estimate the statistical error in the cross validation estimate
\item estimate the dependency of the random POD model on the snapshot. In other words, can further (or other) samples increase the predictivity of the POD model.
\end{itemize}
We give in the following some indications on how to perform these tasks.

\section{Analysis and results}

\subsection{Problem dimensionality away from the cracks}

Given the error map \ref{fig:ErrorPOD16doubleCV}, one can now decide, for a given size of the process zone, of the dimensionality of the problem corresponding to the reduced degrees of freedom. The general problem of identifying the meaningful components in an empirical POD model is a difficult one. It has been addressed in numerous publications (see the reviews and test studies proposed in \cite{jackson1993,cangelosigoriely2007}). Broadly speaking, \cite{cangelosigoriely2007} seems to indicate that all proposed indicators (eg.: elbow tests, tests based on the stability of the eigenvalues of the POD operator like the broken stick or techniques based on bootstrap confidence intervals, and tests based on the rate of predictivity of the model as a function of the number of modes) perform well for a specific range of problems. The author recommends to use several of them in order to avoid relying on an indicator which would be out of its domain of application. We follow this idea. However, our problem differs from that of a classical proper orthogonal decomposition. For a given size $P^r$ of the exclusion zone, the models obtained for different order of the POD approximation are fitted on different sets of data. The corresponding greedy iterates of extractor $\M{\mathbf{E}}^r$ might differ from one order to the next, despite the fact that their ranks are equal. Therefore, the use of tests based on the stability of the eigenvalues of the POD operator is hardly justified. However, we found that tests based on the decrease rate of the predictivity of the model perform well is this context.

The two indicators used in this work are reviewed in \cite{abdiwilliams2010}. The first one identifies the order at which the overfitting behaviour of the empirical POD becomes too large compared to the decrease rate of the cross-validation estimate of the projection error. The number of modes selected using this criterion is \cite{stone1974}:
\begin{equation}
n_\phi^Q = \underset{\tilde{n}_\phi \in \mathcal{Q}_\phi }{\textrm{min}}  \ \left(  \tilde{n}_\phi \right) 
\quad \textrm{where} \quad 
\mathcal{Q}_\phi = \left\{  \tilde{n}_\phi \in \llbracket 1, n_u-1 \rrbracket \ \left| \   \frac{ \bar{\nu}_{| \tilde{n}_\phi+1}^{s,r}  }{ \nu_{| \tilde{n}_\phi }^{s,r}} < 1     \right.     \right\}
\end{equation}
In the above formula, $\bar{\nu}_{| \tilde{n}_\phi}^{s,r}$ is the (double) cross-validation estimate for a restricted POD model of order $\tilde{n}_\phi$, while $\nu_{| \tilde{n}_\phi+1 }^{s,r}$ is the value of $\nu^{s,r}$ (biased estimate of the projection error) for a model of order $\tilde{n}_\phi+1$. It is usually argued that this criterion tends to underestimate the number of modes that should be retained. Indeed, an overfitting behaviour does not necessarily imply that the corresponding modes cannot yield a significant decrease in the projection error. The second criterion used in this work measures the rate of predictivity of the POD model as a function of the number of modes. This rate is compared to the rate at which statistical degrees of freedom are fixed by the tensor approximation of the samples (for more details, see \cite{krzanowski1987}). The number of modes selected using this criterion is:
\begin{equation}
n_\phi^W = \underset{\tilde{n}_\phi \in \mathcal{W}_\phi }{\textrm{min}}  \ \left(  \tilde{n}_\phi \right) 
\quad \textrm{where} \quad 
\mathcal{W}_\phi = \left\{  \tilde{n}_\phi \in \llbracket 1, n_u-1 \rrbracket \ \left| \   \frac{ \left( \bar{\nu}_{| \tilde{n}_\phi}^{s,r} \right)^2 - \left( \bar{\nu}_{| \tilde{n}_\phi+1}^{s,r} \right)^2  }{ \left(  \bar{\nu}_{| \tilde{n}_\phi}^{s,r}  \right)^2 } \, \Delta_{|\tilde{n}_\phi}  < 1     \right.     \right\}
\end{equation}
In the last equation, $\displaystyle \Delta_{|\tilde{n}_\phi} = \frac{(n_u-\tilde{n}_\phi)(n_\theta-\tilde{n}_\phi)}{(n_u-\tilde{n}_\phi)+(n_\theta-\tilde{n}_\phi) -1}$ (the expression of this quantity has been adapted to the version of the POD at use in this paper, which works with samples that have not been centred prior to the spectral analysis). This criterion is usually found to be more conservative than the previous one.

Using an average of both these cross-validation based criteria performs well in our example. For a given size of the process zone, we set $n_\phi$ to the number of modes that corresponds to the average of the cross-validation estimates obtained with a restricted POD of order $n_\phi^Q$ and $n_\phi^W$. The result of our indicator is plotted in figure\ref{fig:ErrorPOD16doubleCV}. Its validity will be commented using qualitative arguments later on.


\subsection{Spatial exclusion zones}

Once the dimensionality of the random process is determined, depending on the size of the excluded spatial domain, one can extract one particular POD model and its associated process zone. We recall the observation that in figure \ref{fig:ErrorPOD16doubleCV}, the cross-validation estimate of the projection error decreases continuously with the rank of the process zone extractor. There is indeed a narrow region where the rate of this trend is higher than anywhere else (between $P^r = 0$ and $P^r = 10 \%$), which corresponds to all random cracks being excluded from the domain of validity of the restricted POD. However, the effect of the cracks is ``seen'' by the structure even far from this narrow region.

Therefore, process zones need to be defined for a given accuracy of the POD model. The choice of one particular model among those represented with the dashed line in figure \ref{fig:ErrorPOD16doubleCV} has to be done according to practical considerations: the selection is a trade off between accuracy and numerical costs for the ``offline'' phase of the reduced order modelling technique (i.e.: the phase where the model will actually be used to reduce CPU time in further computations). A large excluded domain would mean a large part of the spatial degrees of freedom for which no reduced space is available (or only the regularisation proposed in section \ref{sec:RestrictedPOD}, whose associated projection error is potentially large and not controlled in any case). Conversely, a small excluded domain yields inaccurate predictions in the reduced space built for the remaining degrees of freedom. If one knows what the minimum required accuracy is, one can select the model that gives this value, for a maximum size of the domain of validity. Such a model, associated to an error of $10^{-3}$, is shown in figure \ref{fig:ErrorPOD16doubleCV}. However, in the general case, one would have to specify the way the reduced model would be used, and define and objective function that depends on both the speed-up provided by the reduction and the accuracy. Once this objective function is defined, its minimisation over the admissible models (dash line in figure \ref{fig:ErrorPOD16doubleCV}) would lead to the selection of the desired domain of validity. How to use the model will be elaborated on in the discussion section.

\begin{figure}[htb]
       \centering
       \includegraphics[width=0.9 \linewidth]{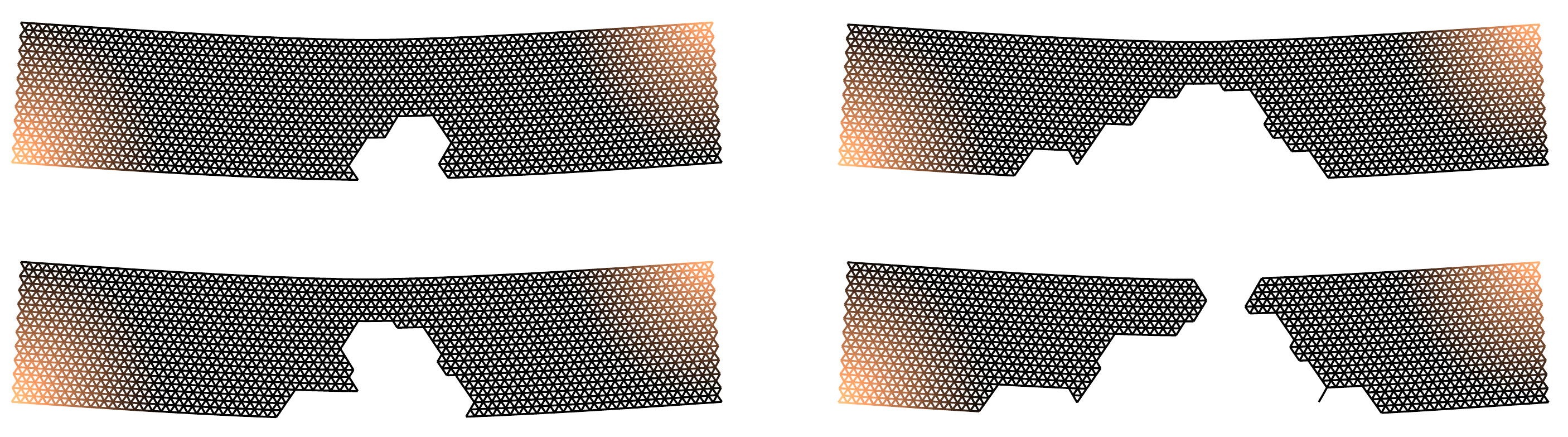}
       \caption{Process zones obtained by using the restricted POD. Increasing the size of the process zone decreases the error of projection of the POD model that is fitted on the remaining spatial degrees of freedom.}
       \label{fig:ProcessZone}
\end{figure}

For purposes of illustration, we show different process zones, associated with increasing values of $P^r$ in figure \ref{fig:ProcessZone} (POD models at circled points of the map in figure \ref{fig:ErrorPOD16doubleCV}). The basis vectors obtained by the Restricted POD are also depicted in figure \ref{fig:ModesFull} for $P^r$ corresponding to a cross-validation estimate of the projection error equal to $10^{-3}$. On can clearly see that the 6 first modes are associated with global deformation of the domain of validity of the reduced model. The 7$^{th}$ (not  retained) can be qualified as ``noise'' engendered by the proximity of the damaged region. This gives us a qualitative confidence in the criterion used to find dimensionality of the problem.

%% file: results.tex
\begin{figure}[htb]
       \centering
       \includegraphics[width=0.9 \linewidth]{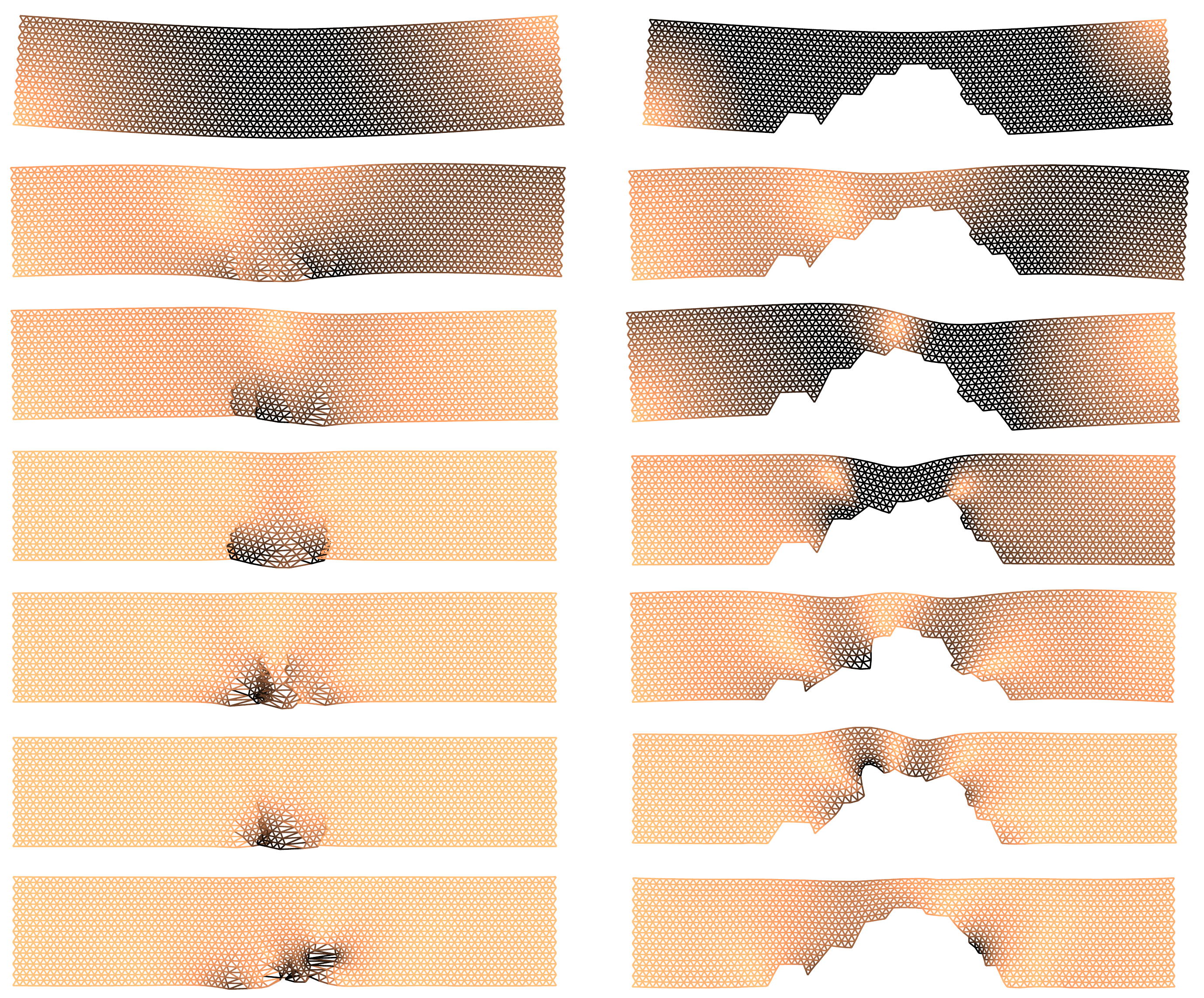}
       \caption{Modes obtained with a restricted POD performed on 16 realisations of the random process (right column), truncated at order 6 (the last vector displayed in this picture is the first mode which is not selected by the criterion used in this paper), and for a size of the process that leads to a projection error of $10^{-3}$ over the complementary spatial domain. In the left column, the POD modes are represented with the regularisation proposed for the spatial components that belong to the process zone.}
       \label{fig:ModesFull}
\end{figure}



\section{Potential applications and improvements}

\subsection{Reduced order modelling and POD-based preconditioners}

From the previous section, it clearly appears that the tools proposed in this paper will have to be adapted to each specific application. In particular, the effort made in the ``offline'' and ``online'' phases of the selected strategy  will have to be balanced in terms of error estimation. We discuss here two different types of potential applications.

\begin{itemize}
\item The restricted-POD as a preconditioner. Reduced-spaces obtained by the POD are a good alternative, or complement, to Krylov-based subspaces for the deflation of iterative solvers, and in particular when the problem is characterised by large changes in the tangent operators \cite{kerfridengosselet2010,carlbergbou-mosleh2011}. Using such a reduced space as a preconditioner is not particularly difficult. As the reduced space is merely used to accelerate linear solvers, separating ``offline'' and ``online'' phases in not necessary. The restricted POD model can be computed, or updated using cheaper algorithms, while considering as samples all the previously computed solutions.
\item The restricted-POD to construct a reduced order model. We look here for a cheap approximation of the (nonlinear) Schur complement of the reducible spatial domain onto the process zone, as proposed in various papers \cite{haryadikapania1998,ammar2011,kerfridenpassieux2011}. In the case, the proposed tool identifies a reduced space in which the Schur complement of the problem with randomness can be projected. However, proper error estimates need to be established. One has here the choice between computing a large number of samples, and using statistical tools to obtain confidence intervals for the projection error, or developing \textit{a priori} or \textit{a posteriori} error estimate that are reliable and cheap enough to be applied ``online'' phase (some of the work related to error estimate of nonlinear reduced order models can be found in \cite{meyermatthies2003,legresleyalonso2003,barraultmaday2004}). The former approach is preferable if interactivity is the goal of the reduced model. However, this is not clear if the purpose is to obtain some statistics of the problem in a minimum amount of time. One then has to balance the effort made to increase the predictivity and confidence of the model in the ``offline'' phase, and the cost of ``online'' error estimates and updating procedures. This is a difficult issue, notably as both statistical error estimate for the POD, and error estimates for general nonlinear reduced order model, are not yet available, which is further discussed later on.
\end{itemize}

\subsection{Statistical extraction and reduced order modelling for general structural heterogeneities}

Let us recall that in the example studied throughout this paper, the elastic properties of the composite material have been assumed constant. Indeed, non-homogeneous elastic properties imply a global lack of correlation. The deformation of the structure within each phase cannot be represented accurately within a low-dimensional subspace. Therefore, a fine-scale correction of a global POD basis would be necessary for each realisation. We do not believe that this is of significant difficulty for localised nonlinearities (for instance in quasi-brittle). One can use the restricted POD proposed in this paper to obtain an empirical model, based on random samples. In the ``online phase'', one can then add to the global basis the residual of the predictions performed with the empirical model, and this for a few of the time steps of the analysis. However, for global nonlinearities (e.g.: large deformations, ductile fracture), this framework might be jeopardised as finding a subspace for the residual of the empirical POD model might require to solve all realisations exactly. One possible avenue for the \textit{a priori} reduced modelling of general heterogeneous structure might be the use of mappings from realisations of the random heterogeneities to a reference configuration.

An other issue to consider in the context of reduced order model of nonlinear problems is that significant speed-up cannot be achieved within a Galerkin framework. One has to devise strategies such as Petrov-Galerkin techniques \cite{barraultmaday2004,ryckelynck2005,astridweiland2008}, or trajectory interpolations \cite{rewienskiwhite2003}. In the former case, the nonlinear behaviour is somehow extrapolated from a point of the domain to its surrounding, hence reducing the costs of numerical integration requires to evaluate the residual of nonlinear problems. Such a methodology, in the context of heterogeneous structures, is far from being obviously applicable. In the second case, the local nonlinear behaviour is interpolated over sampled values of the weighting functions associated with the reduced basis vectors.
In addition, our recent investigations show that the reducibility of a problem is not directly related to the dimensionality of the ensemble of possible solutions, but to the dimensionality of its image by the nonlinear operator of consideration \cite{kerfridengoury2012}. Therefore, this image should be identified and analysed, possibly with the tools proposed in this paper.

\subsection{Evaluation of the sensitivity of the statistical model to the realisations}

As mentioned earlier, the selection of the process zone depends on an objective function which defines the trade-off between the cost and accuracy of the method. Once a threshold of accuracy is selected by the user, the algorithm computes a greedy optimum ($\M{\mathbf{\Phi}}^s, \M{\mathbf{E}}^r$), which minimizes the cost function. Since the optimum depends on the random set of samples $\Theta^s$ and the selection of the complementary sets $\Theta_\phi^s$ and $\Theta_E^s$, it is desirable to determine the sensitivity of the optimum to these realisations.

This could be accomplished by the method of ``Repeated Double Cross Validation'' introduced in \cite{filzomerliebmann2009}. The idea of this method is to repeat the double cross validation $n_{Rep}$ times with different splittings of the set $\Theta^s$ in $\Theta_\phi^s$ and $\Theta_E^s$. This yields $n_{Rep}$ greedy-optimal pairs ($\M{\mathbf{\Phi}}^s, \M{\mathbf{E}}^r$), from which the variability of the outputs can be estimated. The number of repetitions $n_{Rep}$ depends on the complexity of the model studied, values between 10 and 100 have been suggested in the literature \cite{filzomerliebmann2009, koutsoulerisgaser2010}.

In order to analyse the convergence of this method, we need to define a distance function which induces a topology on the space of outputs considered. A possibility for this metric is the \textit{Hausdorff distance}, which gives the distance between two subsets of $\mathbb{R}^n$.
First one defines the directed Hausdorff distance from set $A$ to set $B$ by:
\[ h(A,B) := \max_{a \in A} \left( \min_{b\in B} d(a,b) \right) \]
where $d(a,b)$ is the Euclidean metric between two points $a$ and $b$ in $\mathbb{R}^n$. We note that this distance is, in general, not symmetric, that is $h(A,B)$ is not equal to $h(B,A)$. The undirected Hausdorff distance between the sets $A$ and $B$ is defined by:
\[ H(A,B) := \max \left( h(A,B), h(B,A) \right). \]
The Hausdorff distance is commonly used in object matching and pattern recognition \cite{huttenlocherklanderman1993}. Here it can be used to measure the variance between the process zones obtained by different realisations of the model.

We are particularly interested in ``confidence zones'' for the support of the restriction operator $\M{\mathbf{E}}^r$. These could be easily extracted from the frequency with which each lattice node is included in the process zone or its complement. For example, if $n_{Rep} = 10$ and a particular node is included in the process zone at least 9 times, then we can include it in a 90\% confidence area for the process zone (or 10\% confidence for the complement).

This method has the advantage that the samples used for calculating the projector and not used for the projection error (i.e. the training data and the testing data) are kept separate. This eliminates too optimistic error estimates due to over-fitting and allows us to study the variability of the predicted process zone. If the confidence zones obtained show too much variability, this probably indicates that the number of samples $n_\theta$ is too small and that a larger number of samples is needed.

It is important to realise that these re-sampling techniques always rely on the available realisations to evaluate the variability in the statistics of interest. Therefore, a sufficiently large number of realisations of the random process is required, the term ``sufficiently'' being extremely difficult to quantify as it requires to evaluate the propagation of variabilities through a highly nonlinear structural model.

\subsection{Advanced statistical surrogates and extension to parametric/stochastic problems}

From a conceptual point of view, the method described in this paper is a tailored weighted POD, with an adaptive choice of weights assumed binary. It would however be of interest to consider a transition zone between the process zone, where the lack of correlation is such that no representative reduced space can be identified, and the zone where the data are assumed reducible. Weighting this region will permit to take into account the variability of the solution in this transition zone, and its identification would permit to give identifications about where the region where ``online'' error estimation should be performed.

An other important remark is that we did not make the best solution samples in this work in the sense that we did not segregate spatial correlation in time (within each realisation of the fracture problem), and spatial correlation due to the variability. We can expect that smaller process zones and lower level of projection errors would be obtained by performing the Restricted POD at each time step of the analysis. The idea of performing spectral analysis using a separation of variables is essentially what proposes the proper generalised decomposition (see for instance \cite{ladevezepassieux2009,chevreuilnouy2012}). However, such an extension of the proposed work will require to first define what time is in a quasi-static context, so that random solutions at a given time step share physical similarities (e.g.: crack ``length'').

The method proposed in this paper is illustrated in the context of problems with randomness. The randomness that has been considered in our test case is driven by engineering observations, not from mathematical considerations, which makes the validation of the dependency of the POD model to the particular set of realisations at hand difficult. However, the application of the concept proposed in this paper is not limited to statistics, and could be easily extended to parametric and stochastic problems. In these two contexts, formal or practical error indicators for the POD surrogate would probably be easier to obtain.
